\documentclass[twocolumn]{aastex61}

\usepackage{amsmath, amstext, amsopn, amssymb}
\usepackage{hyperref}

\newcommand{\Msun}{\rm M_{\odot}}
\newcommand{\Lsun}{\rm L_{\odot}}

\newcommand{\mm}{$\mu$m}

\def\gs{\mathrel{\raise0.35ex\hbox{$\scriptstyle >$}\kern-0.6em
\lower0.40ex\hbox{{$\scriptstyle \sim$}}}}
\def\ls{\mathrel{\raise0.35ex\hbox{$\scriptstyle <$}\kern-0.6em
\lower0.40ex\hbox{{$\scriptstyle \sim$}}}}

\shorttitle{Multi-wavelength properties of S2COSMOS sources}
\shortauthors{An et al.}

\begin{document}

\title{Multi-wavelength properties of radio and machine-learning identified counterparts to submillimeter sources in S2COSMOS}

\correspondingauthor{Fang~Xia An}
\email{fangxia@idia.ac.za, fangxiaan@gmail.com}

\author[0000-0001-7943-0166]{Fang\,Xia\,An}
\affiliation{Centre for Extragalactic Astronomy, Department of Physics, Durham University, Durham, DH1 3LE, UK}
\affiliation{Department of Physics and Astronomy, University of the Western Cape, and the Inter-University Institute for Data Intensive Astronomy, Robert Sobukwe Road, 7535 Bellville, Cape Town, South Africa}
\affiliation{Purple Mountain Observatory, Chinese Academy of Sciences, 10 Yuanhua Road, Nanjing 210034, China}

\author{J.\,M.\,Simpson}
\affiliation{Academia Sinica Institute of Astronomy and Astrophysics, No.\,1, Section 4, Roosevelt Rd., Taipei 10617, Taiwan}
\affiliation{Centre for Extragalactic Astronomy, Department of Physics, Durham University, Durham, DH1 3LE, UK}

\author{Ian\,Smail}
\affiliation{Centre for Extragalactic Astronomy, Department of Physics, Durham University, Durham, DH1 3LE, UK}

\author{A.\,M.\ Swinbank}
\affiliation{Centre for Extragalactic Astronomy, Department of Physics, Durham University, Durham, DH1 3LE, UK}

\author{Cong\,Ma}
\affiliation{South African Radio Astronomy Observatory, 2 Fir Street, Observatory 7925, Western Cape, South Africa}
\affiliation{Department of Mathematics and Applied Mathematics, University of Cape Town, Cross Campus Road, Rondebosch 7700, Western Cape, South Africa}
\affiliation{African Institute for Mathematical Sciences, 6--8 Melrose Road, Muizenberg 7945, Western Cape, South Africa}

\author{Daizhong\,Liu }
\affiliation{Max Planck Institute for Astronomy, K{\"o}nigstuhl 17, D-69117 Heidelberg, Germany}

\author{P.\,Lang}
\affiliation{Max Planck Institute for Astronomy, K{\"o}nigstuhl 17, D-69117 Heidelberg, Germany}

\author{E.\,Schinnerer}
\affiliation{Max Planck Institute for Astronomy, K{\"o}nigstuhl 17, D-69117 Heidelberg, Germany}

\author{A.\,Karim}
\affiliation{Argelander-Institut f{\"u}r Astronomie, Universit{\"a}t Bonn, Auf dem H{\"u}gel 71, D-53121 Bonn, Germany}

\author{B.\,Magnelli}
\affiliation{Argelander-Institut f{\"u}r Astronomie, Universit{\"a}t Bonn, Auf dem H{\"u}gel 71, D-53121 Bonn, Germany}

\author{S.\,Leslie}
\affiliation{Max Planck Institute for Astronomy, K{\"o}nigstuhl 17, D-69117 Heidelberg, Germany}

\author{F.\,Bertoldi}
\affiliation{Argelander-Institut f{\"u}r Astronomie, Universit{\"a}t Bonn, Auf dem H{\"u}gel 71, D-53121 Bonn, Germany}

\author{Chian-Chou\,Chen}
\affiliation{European Southern Observatory, Karl Schwarzschild Strasse 2, Garching, Germany}

\author{J.\,E.\,Geach}
\affiliation{Centre for Astrophysics Research, School of Physics, Astronomy and Mathematics, University of Hertfordshire, Hatfield AL10 9AB, UK}

\author{Y.\,Matsuda}
\affiliation{National Astronomical Observatory of Japan, 2-21-1 Osawa, Mitaka, Tokyo, 181-8588, Japan}
\affiliation{The Graduate University for Advanced Studies (SOKENDAI), Osawa, Mitaka, Tokyo, 181-8588, Japan}

\author{S.\,M.\,Stach}
\affiliation{Centre for Extragalactic Astronomy, Department of Physics, Durham University, Durham, DH1 3LE, UK}

\author{J.\,L.\,Wardlow}
\affiliation{Physics Department, Lancaster University, Lancaster LA1 4YB, UK}

\author{B.\,Gullberg}
\affiliation{Centre for Extragalactic Astronomy, Department of Physics, Durham University, Durham, DH1 3LE, UK}

\author{R.\,J.\,Ivison}
\affiliation{European Southern Observatory, Karl Schwarzschild Strasse 2, Garching, Germany}
\affiliation{Institute for Astronomy, University of Edinburgh, Royal Observatory, Blackford Hill, Edinburgh EH9 3HJ, UK}

\author{Y.\,Ao}
\affiliation{Purple Mountain Observatory, Chinese Academy of Sciences, 10 Yuanhua Road, Nanjing 210034, China}

\author{R.\,T.\,Coogan}
\affiliation{Astronomy Centre, Department of Physics and Astronomy, University of Sussex, Brighton BN1 9QH, UK}
\author{A.\,P.\,Thomson}
\affiliation{The University of Manchester, Oxford Road, Manchester M13 9PL, UK}

\author{S.\,C.\,Chapman}
\affiliation{Department of Physics and Atmospheric Science, Dalhousie University, Halifax, NS B3H 3J5, Canada}

\author{R.\,Wang}
\affiliation{Kavli Institute for Astronomy and Astrophysics, Peking University, Beijing 100871, People’s Republic of China}

\author{Wei-Hao\,Wang}
\affiliation{Academia Sinica Institute of Astronomy and Astrophysics, No.\,1, Section 4, Roosevelt Rd., Taipei 10617, Taiwan}

\author{Y.\,Yang}
\affiliation{Korea Astronomy and Space Science Institute, 776 Daedeokdae-ro, Yuseong-gu, Daejeon 34055, Republic of Korea}

\author{R.\,Asquith}
\affiliation{School of Physics \& Astronomy, University of Nottingham, Nottingham NG7 2RD, UK}

\author{N.\,Bourne}
\affiliation{Institute for Astronomy, University of Edinburgh, Royal Observatory, Blackford Hill, Edinburgh EH9 3HJ, UK}

\author{K.\,Coppin}
\affiliation{Centre for Astrophysics Research, Science and Technology Research Institute, University of Hertfordshire, Hatfield AL10 9AB, UK}

\author{N.\,K.\,Hine}
\affiliation{Centre for Astrophysics Research, Science and Technology Research Institute, University of Hertfordshire, Hatfield AL10 9AB, UK}

\author{L.\,C.\,Ho}
\affiliation{Kavli Institute for Astronomy and Astrophysics, Peking University, Beijing 100871, People’s Republic of China}
\affiliation{Department of Astronomy, Peking University, Beijing, 100087, People’s Republic of China}

\author{H.\,S.\, Hwang}
\affiliation{Korea Astronomy and Space Science Institute, 776 Daedeokdae-ro, Yuseong-gu, Daejeon 34055, Republic of Korea}

\author{Y.\,Kato}
\affiliation{National Astronomical Observatory of Japan, 2-21-1 Osawa, Mitaka, Tokyo, 181-8588, Japan}

\author{K.\ Lacaille}
\affiliation{Department of Physics and Astronomy, McMaster University, Hamilton, ON L8S 4M1, Canada}

\author{A.\,J.\,R.\,Lewis}
\affiliation{Institute for Astronomy, University of Edinburgh, Royal Observatory, Blackford Hill, Edinburgh EH9 3HJ, UK}

\author{I.\,Oteo}
\affiliation{European Southern Observatory, Karl Schwarzschild Strasse 2, Garching, Germany}
\affiliation{Institute for Astronomy, University of Edinburgh, Royal Observatory, Blackford Hill, Edinburgh EH9 3HJ, UK}

\author{J.\,Scholtz}
\affiliation{Centre for Extragalactic Astronomy, Department of Physics, Durham University, Durham, DH1 3LE, UK}

\author{M.\,Sawicki}
\affiliation{Department of Astronomy and Physics, Saint Marys University, Halifax, NS B3H 3C3, Canada}

\author{D.\,Smith}
\affiliation{Centre for Astrophysics Research, Science and Technology Research Institute, University of Hertfordshire, Hatfield AL10 9AB, UK}

\begin{abstract}
We identify multi-wavelength counterparts to 1,147 submillimeter sources from the S2COSMOS SCUBA-2 survey of the COSMOS field by employing a recently developed radio$+$machine-learning method trained on a large sample of ALMA-identified submillimeter galaxies (SMGs), including 260 SMGs identified in the AS2COSMOS pilot survey. In total, we identify 1,222 optical/near-infrared(NIR)/radio counterparts to the 897 S2COSMOS submillimeter sources with $S_{\rm 850}>1.6$\,mJy, yielding an overall identification rate of ($78\pm9$)\%. 
We find that ($22\pm5$)\% of S2COSMOS sources have multiple identified counterparts. We estimate that roughly 27\% of these multiple counterparts within the same SCUBA-2 error circles very likely arise from physically associated galaxies rather than line-of-sight projections by chance. The photometric redshift of our radio$+$machine-learning identified SMGs ranges from $z=0.2$ to 5.7 and peaks at $z=2.3\pm0.1$. The AGN fraction of  our sample is ($19\pm4$)\%, which is consistent with that of ALMA SMGs in the literature. Comparing with radio/NIR-detected field galaxy population in the COSMOS field, our radio$+$machine-learning identified counterparts of SMGs have the highest star-formation rates and stellar masses. These characteristics suggest that our identified counterparts of S2COSMOS sources are a representative sample of SMGs at $z\ls3$. We employ our machine-learning technique to the whole COSMOS field and identified 6,877  potential SMGs, most of which are expected to have submillimeter emission fainter than the confusion limit of our S2COSMOS  surveys ($S_{850\,\mu m} \ls 1.5$\,mJy). We study the clustering properties of SMGs based on this statistically large sample, finding that they reside in high-mass dark matter halos ($(1.2\pm0.3)\times10^{13}\,h^{-1}\,\Msun$), which suggests that SMGs may be the progenitors of massive ellipticals we see in the local Universe. 
\end{abstract}

\keywords{Observational astronomy --- Submillimeter astronomy --- Galaxy formation --- Galaxy evolution --- Galaxies: High-redshift galaxies --- Galaxies: Starburst galaxies --- Clustering}

\section{Introduction} \label{sec:intro}
Understanding how galaxies form in the early Universe and their subsquence evolution through cosmic time is a fundamental goal of modern astrophysics. The discovery of a population of dusty galaxies at high redshifts at far-infrared (FIR) and millimeter/submillimeter wavelengths have a profound impact on our studying of galaxies formation and evolution \citep[e.g.,][and see \cite{Casey14} for a review]{Smail97,Barger98,Hughes98,Scott02,Scott12,Coppin06,Weiss09,Yamamura10, Ikarashi11,Clements11,Geach17,Simpson19}. The brighter examples of these FIR/submillimeter sources have infrared luminosities of $L_{\rm IR}$ $\ge 10^{12}\,\Lsun$, which is comparable to the local ultra-luminous infrared galaxies (ULIRGs). Although such infrared luminous galaxies are rare in the local Universe, their spatial density increases rapidly with look-back time and appears to peak at $z\sim$2--3 \citep[e.g.,][U. Dudzevi{\^c}i{\=u}t{\.e} et al.\ 2019, in preparation]{Barger99,Chapman05, Smolcic12, Yun12, Simpson14,Swinbank14, Chen16}. Therefore, these FIR/submillimeter luminous sources host the most intense star formation in the early Universe with star-formation rates (SFRs) of $\ge\,10^{2}$--$10^{3}\,\Msun$\,yr$^{-1}$, which would enable them to form the stellar mass of massive galaxies ($M\ge\,10^{11}\,\Msun$) within $\sim$ 100\,Myr \citep[e.g.,][]{Chapman05, Bothwell13,Casey14}. These characteristics make these FIR/submillimeter bright sources a key element to constrain models of galaxy formation and evolution. 

Thanks to the strong negative $k$-correction in the submillimeter/millimeter observational wavebands, we can detect these ultra-luminous infrared galaxies at high redshift ($z\sim$1--6) with a nearly constant sensitivity in terms of dust mass or far-infrared luminosity (although the latter is sensitive to dust temperature). The majority of bright submillimeter sources have been detected in panoramic, ground-based single-dish submillimeter surveys in the past two decades \citep[e.g.,][]{Scott02,Scott12,Coppin06,Weiss09,Ikarashi11,Wang17, Geach17,Simpson19}. However, the typical angular resolution of these ground-based single-dish submillimeter surveys is $\sim10\arcsec$--20$\arcsec$ at 450--1100\,\mm. This coarse resolution made it very difficult to identify the multi-wavelength counterparts of these submillimeter sources and thus is the major challenge for exploiting these panoramic single-dish submillimeter surveys. The indirect tracers of FIR/submillimeter emission, such as radio, 24\,\mm, or mid-infrared (MIR) properties are traditionally used to identify counterparts to single-dish submillimeter sources \citep[e.g.,][]{Ivison98, Smail02, Pope06, Ivison07, Barger12, Michalowski12, Cowie17}. Unfortunately, the completeness of these identifications is typical $\ls$ 50\% because of the lack of negative $k$-correction and the limited observational depth in radio and MIR bands \citep[e.g.,][although see \citealp{Lindner11}]{Hodge13, Chen16}. The completeness is defined as the number of recovered candidate submillimeter galaxies (SMGs) versus the total number of SMGs in the submillimeter survey.

Recently, the field has advanced considerably as a result of interferometric observations at submillimeter/millimeter wavelength undertaken with the Submillimeter Array (SMA), IRAM's Plateau de Bure Interferometer (PdBI) and Northern Extended Millimetre Array (NOEMA), especially the Atacama Large Millimeter/submillimeter Array (ALMA). These facilities can reach arcsecond/sub-arcsecond positional precision of SMGs, which significantly improved our understanding of these high redshift, dusty starburst galaxies \citep[e.g.,][]{Frayer98,Gear00,Tacconi06, Genzel10, Smolcic12, Hodge13, Swinbank14, Swinbank15, Thomson14,Aravena16, Walter16, Danielson17, Dunlop17, Simpson17, Wardlow17, Gullberg18,Cooke18, Stach18,Stach19}. However, for the large single-dish submillimeter surveys, e.g., the SCUBA-2 Cosmology Legacy Survey \citep[S2CLS;][]{Geach17} and the S2COSMOS survey \citep{Simpson19}, the high-resolution interferometric follow-up is  still challenging to complete.

Previous work has tried to take advantages of both single-dish (efficient large area surveys) and interferometric (high angular resolution) submillimeter observations to provide a large sample of SMGs with precisely identified multi-wavelength counterparts, which is necessary for investigating the statistical properties, such as spatial clustering, of SMGs \citep[e.g.,][]{Hickox12, Chen16, Wilkinson17}. Galaxy clustering is a key measurement that constrains theoretical models of galaxy formation and evolution, since it provides information of the mass of the halos in which the galaxies reside \citep[e.g.,][]{Mo96, Mo02, Cooray02}. 

Galaxies that follow similar evolutionary tracks are expected to reside in halos with similar masses across cosmic time. Because of their intensively star-forming, massive, and high-redshift nature, SMGs have been suggested to be the progenitors of compact quiescent galaxies at $z\sim1$--2 and subsequently local massive ellipticals \citep[e.g.,][]{Hughes98, Eales99, Swinbank06, Targett11, Simpson14, Toft14, Wang19}. This scenario can be tested by comparing the spatial clustering of SMGs and other massive galaxy populations at low redshift or in the local Universe. 

However, because of the coarse angular resolution of single-dish submillimeter surveys and the small survey area of interferometric observations, measurements of SMG clustering have suffered from large uncertainties \citep[e.g.,][]{Weiss09,Williams11, Hickox12, Wilkinson17}. In addition, the previous studies only include brighter SMGs ($S_{\rm 850\,\mu m}\gs$\,2--3\,mJy), despite that faint SMGs are necessary for a more complete picture of SMG spatial distribution in general. Indeed, it has been suggested that the mass of fainter SMGs' host halos may be comparable with the hosts of the brighter SMGs and that the fainter SMGs may contribute $\sim$80\% of the $S_{\rm 850\,\mu m}$ extragalactic background light \citep[e.g.,][]{Cowie02, Chen16b}. However, it is impossible to detect faint SMGs through blank-field single-dish submillimeter surveys if the submillimeter emission falls below the corresponding confusion limit \citep{Jauncey68}. Although ALMA observations can detect the faintest SMGs with $S_{\rm 850\,\mu m} \ls 1$\,mJy, their survey area is very limited \citep{Franco18,Umehata18}. By utilising ALMA survey, \cite{Chen16,Chen16b} developed an optical-infrared triple color-color (OIRTC) technique to select faint SMGs in the UKIDSS-UDS field and measure the clustering strength of SMGs. The main limitation in \cite{Chen16b} was the small sample size of training SMGs and the moderate survey area of the UDS field ($\sim$1\,degree$^{2}$), which in combination caused large uncertainties in clustering measurements for both SMGs and comparison samples, especially at high redshift.

To exploit deep, wide-field single-dish submillimeter surveys \citep{Geach17,Simpson19} and obtain a statistically larger and more robust sample of counterparts to SMGs, more  advanced techniques for counterpart identification are required. By utilizing a large sample of ALMA identified SMGs from the ALMA follow-up of the S2CLS submillimeter sources in the UDS field \citep[AS2UDS;][]{Stach18,Stach19} as a training set, we developed a machine-learning method to identify multi-wavelength counterparts of single-dish submillimeter sources in \cite{An18} (hereafter An18), and it was supplemented by the use of radio emission as an indirect tracer of submillimeter emission \citep[e.g.,][]{Ivison02, Ivison07, Biggs08,Thomson14}. The robustness of our method is confirmed by a series of self-tests and independent tests as shown in An18. 

In this work, we employ the same radio$+$machine-learning method developed in An18 to our new SCUBA-2 submillimeter survey in the COSMOS field \citep[S2COSMOS;][]{Simpson19} to obtain a large sample of SMGs across a wide field with reliably identified counterparts and investigate their physical and evolutionary properties. The observations of our test sample and training sets, including ALMA identified SMGs from AS2COSMOS pilot survey (J. M. Simpson et al.\,2019 in preparation), as well as the ancillary data in the COSMOS field are introduced in \S\ref{s:observation}. We present our analyses of radio and machine-learning identification of multi-wavelength counterparts to S2COSMOS sources in \S\ref{s:identification}. We give our results and discussions of the multi-wavelength and clustering properties of SMGs in \S\ref{s:main_results}. The main conclusions of this work are given in \S\ref{s:conclusion}. Throughout this paper, we adopt the AB magnitude system \citep{Oke74} and assume a flat $\Lambda$CDM cosmological model with parameters fixed at the $Planck$ 2015 best-fit values, namely, the Hubble constant $H_0 = 67.27$\,km\,s$^{-1}$\,Mpc$^{-1}$, matter density parameter $\Omega_{\rm m} = 0.32$, and cosmological constant $\Omega_{\Lambda} = 0.68$ \citep{Planck16}.

\section{Observations} \label{s:observation}
\subsection{S2COSMOS}
The 850\,\mm\ SCUBA-2 COSMOS survey (S2COSMOS) was carried out with the East Asian Observatory’s James Clerk Maxwell Telescope (JCMT) between Jan.\ 2016 and Jun.\ 2017. We provide a brief overview here, and the full details of observations, data reduction, and catalog are described in \cite{Simpson19}. For S2COSMOS, we adopted a similar observing strategy to S2CLS \citep{Geach17} since the partially completed S2CLS map of COSMOS is incorporated into our S2COSMOS survey. Specifically, we first use four PONG-2700 scans, which provide a uniform coverage over a circular region with diameter of 45$\arcmin$ located equidistant from the centre of the field to map the full 2\,degree$^{2}$ COSMOS field. To reduce the inhomogeneous sensitivity caused by the scan overlap, we adopt a smaller scan pattern, PONG-1800, with scan diameter of 30$\arcmin$ to obtain observations in the four corners of the COSMOS field \citep[see Fig~1 in][]{Simpson19}. The total exposure time is 223\,hr with the PONG-2700 and PONG-1800 scans in a ratio of five-to-one. Combining with the SCUBA-2 archival imaging data at 850\,\mm, which are mostly from S2CLS \citep{Geach17}, in total, we consider a 640\,hr wide-field 850\,\mm\ map of the COSMOS field.

As described in \cite{Simpson19}, the S2COSMOS data were reduced by using the process described in \cite{Chapin13} with the Dynamical Iterative Map Maker ({\sc dimm}) within the Sub-Millimeter Common User Facility ({\sc smurf}), which is provided as part of the {\sc starlink} software suite. We refer the reader to \cite{Simpson19} for the details of the data reduction procedures.  

The instrumental sensitivity varies across the final map. In the centre of the image, where the four PONG-2700 scan patterns overlap, the lowest noise reaches $\sigma_{850\rm\,\mu m}=0.5$\,mJy\,beam$^{-1}$, while in the outer regions the instrumental noise increases to $\sigma_{850\rm\,\mu m}\le5$\,mJy. Therefore, \cite{Simpson19} defined a 1.6\,degree$^{2}$ region matching the {\it HST}/ACS footprint as the S2COSMOS {\sc main} survey region with median noise level of $\sigma_{850\rm\,\mu m}=1.2$\,mJy\,beam$^{-1}$ and an additional surrounding 1\,degree$^{2}$ supplementary ({\sc supp}) survey region with a median 1-$\sigma$ instrumental sensitivity of 1.7\,mJy\,beam$^{-1}$. \cite{Simpson19} present catalogs of the sources detection within these {\sc main} and {\sc supp} regions. The empirical point spread function (PSF) of S2COSMOS survey is obtained by stacking bright, isolated sources and has an FWHM of 14$\farcs$8. In total, 1020 and 127 submillimeter sources are detected at a significance level of $>4\,\sigma$ and $>4.3\,\sigma$ in the {\sc main} and {\sc supp} regions, respectively, corresponding to a uniform false detection rates of 2\% \citep{Simpson19}. In this work, we use the whole sample of 1,147 sources in our analysis. 

\subsection{Training set: ALMA observations in the COSMOS field} \label{sec:obs_ts}
The use of a larger training sample ensures better performance of machine-learning methods. Hence we prefer to use the largest available sample of ALMA-identified SMGs in the COSMOS field, supplemented by large ALMA samples in other fields, e.g., AS2UDS \citep{Stach19}, as a training set for identifying multi-wavelength counterparts of our S2COSMOS sources. 

\subsubsection{AS2COSMOS}
We have completed an ALMA Cycle 4 pilot study of the brightest 160 single-dish submillimeter sources from S2COSMOS (Project ID: 2016.1.00463.S). The ALMA follow-up observations were taken in Band 7 (870\,\mm) in April and May 2018. The ALMA primary beam diameter at this frequency is 17$\farcs$3, which encompasses the area of the SCUBA-2 beam. In addition, there are 24 archival ALMA maps at Band 7 corresponded to our S2COSMOS sources. Therefore, our full AS2COSMOS sample includes 184 ALMA maps with median sensitivity $\sigma_{\rm 870\,\mu m}=0.2$\,mJy\,beam$^{-1}$. The median synthesised beams of the 184 ALMA maps is 0$\farcs8\times0\farcs79$. Full details of the observations are presented in J. M. Simpson et al. (in preparation).  

The ALMA data were reduced using the the {\sc Common Astronomy Software Application} \cite[CASA,][]{McMullin07} v4.2.2-5.1.1. The data reduction procedures and source detection method are similar to that of the ALMA-SCUBA-2 Ultra Deep Survey \citep[AS2UDS;][]{Stach18, Stach19}. In the 184 ALMA maps, 260 ALMA SMGs were detected at a peak signal-to-noise ratio (SNR) $>4.8$ or a 1$\farcs$2-diameter aperture SNR $>4.9$. 

\subsubsection{A$^{3}$COSMOS}

We use additional ALMA archival data in the COSMOS field to construct our training set. All the publicly available ALMA archive data in the COSMOS field were processed, imaged and analysed in the on-going ALMA archive mining project A$^{3}$COSMOS \citep{dLiu19}\footnote{https://sites.google.com/view/a3cosmos}. The latest version of the A$^3$COSMOS data set contains 1,909 ALMA pointings, leading to 1,134 robust ALMA detections with SNR $\gtrsim5.4$, corresponding to a spurious fraction of 8\% and a completeness\, $>90\%$ \citep{dLiu19}.  To limit the complications that arise from different observational frequencies, we only use those continuum observations with frequencies close to the central frequency of the SCUBA-2 filter transmission ($800<\lambda_{c}<1,200$\,\mm). We convert the flux density of all A$^{3}$COSMOS SMGs to $S_{870\,\mu m}$ by adopting the ratios given in \cite{Fujimoto16} and limited our training set to $S_{870\,\mu m}>1$\,mJy. We also remove very shallow observations by limiting the root-mean-square (RMS) noise of the map to $\sigma_{\rm 870\,\mu m}\le0.25$\,mJy. In total, 984 A$^{3}$COSMOS SMGs meet these requirements and are used in constructing the training set for identifying counterparts to S2COSMOS submillimeter sources.

\subsection{Additional Multi-wavelength Observations/Catalogs}
The COSMOS field is one of the largest extragalactic fields with a rich ancillary dataset. Here we give a short summary of the data that we use in our analysis. 

\subsubsection{VLA-3\,GHz radio maps}
The radio data we use in this work are from the VLA-COSMOS 3\,GHz Large Project \citep{Smolcic17}, which were taken with the Karl G. Jansky Very Large Array (VLA) at 3\,GHz. The reduced continuum data and source catalog have been released by \cite{Smolcic17}. In summary, the median RMS of the final reduced data reaches 2.3\,$\mu$Jy (equivalent to $\simeq$\,4\,$\mu$Jy RMS at 1.4\,GHz) over the 2\,degree$^{2}$ COSMOS field with an angular resolution of 0$\farcs$75. \cite{Smolcic17} detected 10,830 3\,GHz radio sources at $\ge5\,\sigma$, which are used to identify radio counterparts of S2COSMOS submillimeter sources in our work. 

\subsubsection{Optical/NIR/FIR catalogs}\label{s:ONFcatalog}
The COSMOS2015 catalog \citep{Laigle16} from UltraVISTA-DR2 surveys are used to identify optical/NIR counterparts for both ALMA SMGs and S2COSMOS submillimeter sources and are then used to construct a training set and a test sample for the machine-learning analyses. In An18, we found that photometric redshift, absolute $H$-band magnitude and NIR colors have the greatest diagnostic power to differentiate SMGs from non-SMGs. Since the COSMOS field has deep $z^{++}$ data, which are also used to detect sources in the region that lie outside of the coverage of the $K_{\rm s}$ band \citep{Laigle16}, we use the $z^{++}$, $J$, $K_{\rm s}$, IRAC 3.6\,\mm\ and 4.5\,\mm-band photometries from the COSMOS2015 catalog in our machine-learning analyses. 

The 3\,$\sigma$ depth of $z^{++}$ is 25.9\,mag in a 3$\arcsec$ diameter aperture. The UltraVISTA-DR2 has a $J$ and $K_{\rm s}$-band observations reaching a 3\,$\sigma$ depth of $J=24.7$\,mag and  $K_{\rm s}=24.0$\,mag in a region of 1.7\,degree$^{2}$ and a $J=24.9$\,mag and $K_{\rm s}=24.7$\,mag in the four ultra-deep stripes, which cover an area of 0.62\,degree$^{2}$. The IRAC 3.6\,\mm\ and 4.5\,\mm\ observations have 3\,$\sigma$ depth of 25.5\,mag within a 3$\arcsec$ diameter aperture. 
\cite{Laigle16} used the total flux, which is estimated from the corrected 3\,$\arcsec$ aperture flux, to fit the SEDs from near-ultraviolet to NIR for their NIR-detected sources. They produced the probability distribution function (PDF) of photometric redshift for each galaxy by matching the observed SED to a set of galaxy templates at a redshift grid with a step of 0.01 and range of $0<z<6$ through minimizing the $\chi^{2}$.  The median of this distribution was determined as the photometric redshift of galaxy in COSMOS2015 catalog. The absolute $H$-band magnitudes given in the catalog was estimated from the best-fitting SEDs. We refer the reader to \cite{Laigle16} for the details.

The other photometric catalog used in this work is a ``super-deblended'' FIR to (sub)millimeter photometric catalog from \cite{Jin18}. Using the position of $K_{\rm s}$-band or radio (in the case of $K_{\rm s}$-band non-detection) sources, \cite{Jin18} adopt a ``super-deblended'' method developed by \cite{Liu18} to ``deblend'' the FIR to (sub)millimeter photometry of sources from {\it Spitzer} \citep{Le09}, {\it Herschel} \citep{Oliver10, Lutz11, Bethermin12}, SCUBA2 \citep{Cowie17, Geach17}, AzTEC \citep{Aretxaga11}, MAMBO \citep{Bertoldi07} and VLA (1.4\,GHz and 3\,GHz) surveys \citep{Schinnerer10, Smolcic17} in the COSMOS field. \cite{Jin18} estimated the SFR of NIR or radio-selected galaxies by integrating 8-1000\,\mm\ infrared luminosities derived from FIR$+$millimeter SED fitting. In our work, we use these estimates of SFR to investigate the star-formation efficiency of our identified counterparts of SMGs in the COSMOS field.

\section{Analysis}\label{s:identification}

In a similar manner to An18, in order to maximize the completeness, we combine the radio and machine-learning methods to identify the multi-wavelength counterparts of single-dish submillimeter sources.

\subsection{Radio identification} \label{s:identification_radio}
Radio synchrotron emission has been proven to be a useful tracer of obscured star formation, since it is powered by supernova remnants of massive stars. Therefore, radio identification is a traditional method to search for counterparts of submillimeter sources \citep[e.g.,][]{Ivison02}. Instead of considering all radio sources within the SCUBA-2 error circles as potential counterparts, we use the corrected-Poissonian probability, $p$-value \citep{Downes86,Dunlop89} to calculate the probabilistic association of radio sources to single-dish submillimeter sources. In An18, we confirmed that the adoption of  $p\le0.065$ increases the precision of radio identification from 64\% to 70\%. In this work, we adopt this limit in the radio identification.

%
%
\begin{figure}[!t]
\centering
\includegraphics[width=0.46\textwidth]{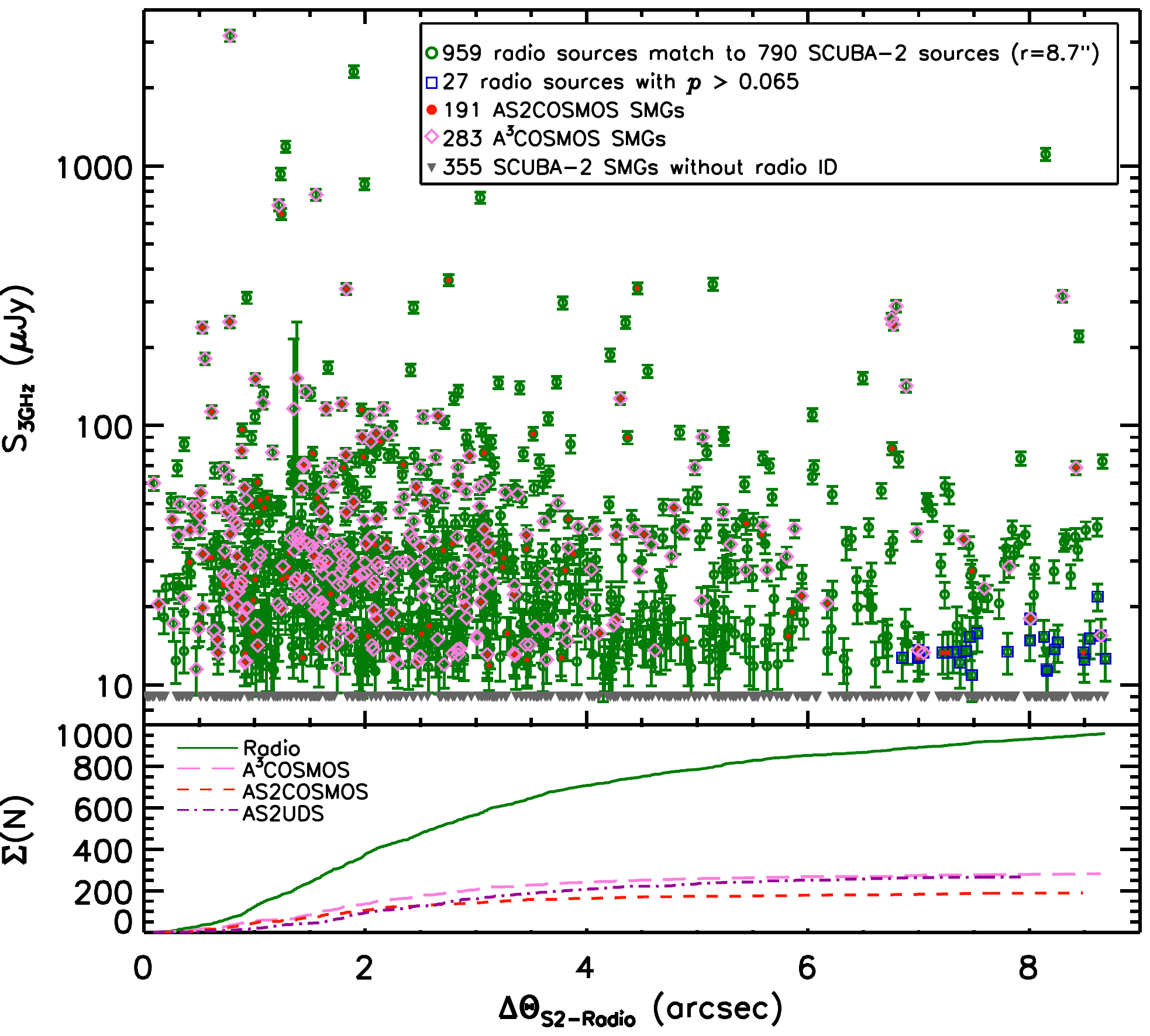}
\caption{{\it Top:} The radio flux densities for all radio sources within the error circles ($r=8\farcs7$) of the SCUBA-2-detected submillimeter sources as a function of the offset of these radio sources from the SCUBA-2 single-dish source. For the 1,147 S2COSMOS submillimeter sources, 1,145 of them are covered by the VLA 3\,GHz radio map. In addition, 959 radio sources lie within 790 SCUBA-2 error circles. We also mark the radio counterparts of ALMA SMGs from our pilot AS2COSMOS survey and from the ALMA archive (A$^{3}$COSMOS). We use $p\le$ 0.065 as a cut of ``robust" radio identification according to our previous work \citep{An18}. From the test of AS2UDS ALMA SMGs in the UDS field, we expect that the precision of radio identification is $\sim$70\%. Among the 959 radio sources, 27 have $p>$0.065 and are removed from our identification catalog. Therefore, 932 counterparts of SCUBA-2 sources in the COSMOS field are identified by the radio data alone. There are also 355 SCUBA-2 sources that do not have radio counterparts within $8\farcs7$. {\it Bottom:} The cumulative number of radio sources and ALMA SMGs as a function of their offset from the SCUBA-2 sources. We also plot the results of AS2UDS SMGs for comparison. In both fields, the number of ALMA SMGs converges at $\Delta \theta>3''$ while the number of radio sources increases gradually, suggesting that these include some associated companions to the submillimeter sources.}
\label{f:radio.eps}
\end{figure}

The VLA 3\,GHz radio observations cover all of the S2COSMOS sources, except for the two northernmost ones. In this work, we define the error circle of a SCUBA-2 source as $r=8\farcs7$, which is the full width at half maximum (FWHM) of the primary beam of ALMA Band 7, since the precision and recall of radio and machine-learning method are all estimated based on the ALMA SMGs in AS2UDS. Therefore, the performance of the radio and machine-learning identification can only be tested within the ALMA primary beam. {\it Precision} is defined as the ratio between the number of correctly identified SMGs and the total number of predicted SMGs using the radio or machine-learning classification. {\it Recall} is the number of corrected classification versus the total number of ALMA SMGs.  In Figure~\ref{f:radio.eps}, we show that there are 959 $\ge 5\,\sigma$ VLA 3\,GHz radio sources within 790 SCUBA-2 error circles. Among them, 932 have $p\le0.065$ and thus are identified as the likely counterparts of S2COSMOS sources. According to the results for AS2UDS in An18, the precision of radio identification is around 70\%. The radio imaging in the COSMOS field is $\sim2\times$ deeper than that in the UDS field, therefore, the completeness of radio identification in this work should be higher than that of AS2UDS (39\%). However, as shown in Figure~\ref{f:radio.eps}, there are 355 S2COSMOS sources without radio identifications. This may be caused by the fact that the radio observation do not benefit from a negative $k$-correction, which means even deep radio observation will miss higher-redshift SMGs. Therefore, other methods of identifying counterparts of S2COSMOS sources are necessary to improve the completeness.

%
%
\begin{figure}[!t]
\centering
\includegraphics[width=0.46\textwidth]{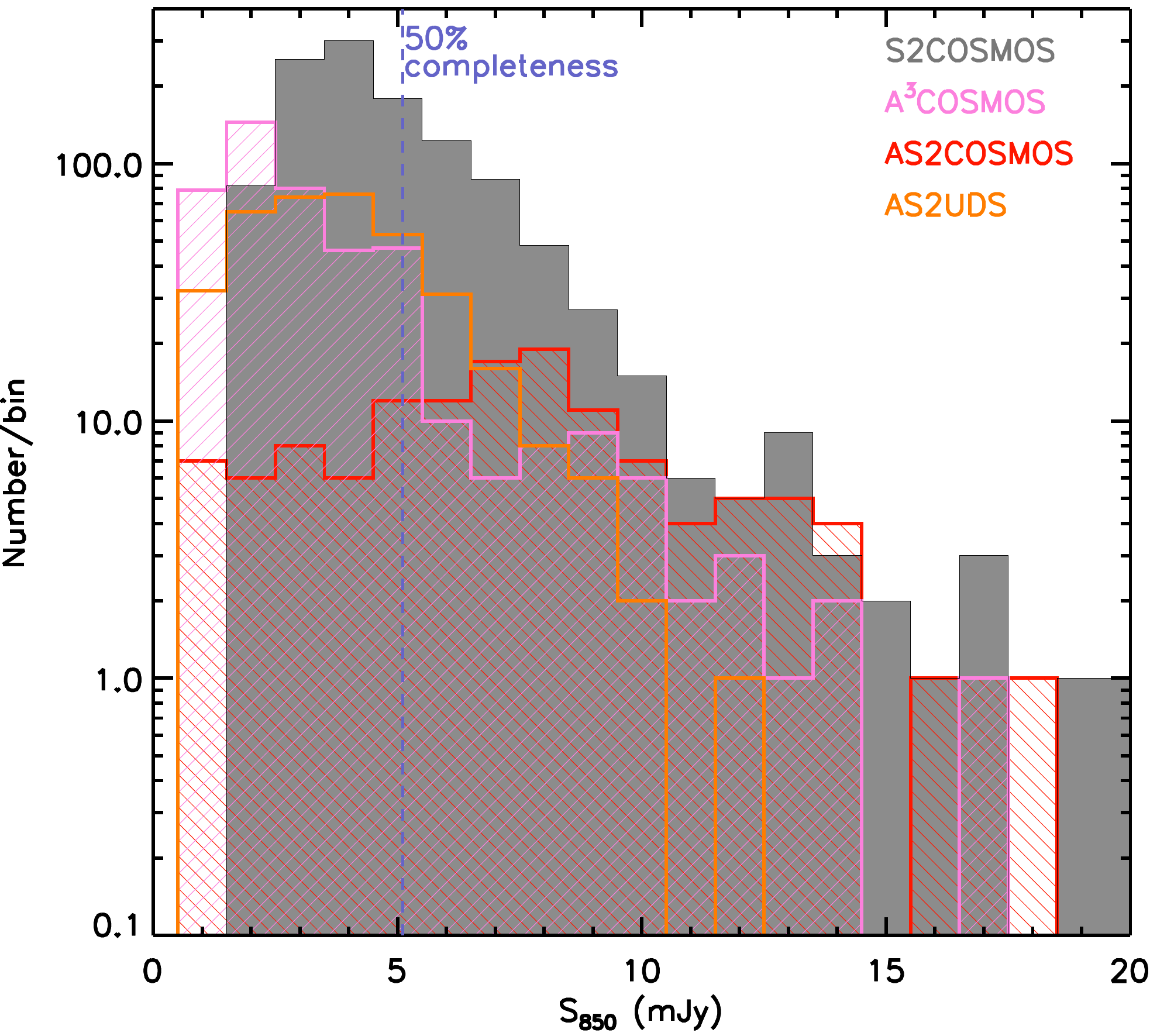}
\caption{Flux density distributions of S2COSMOS submillimeter sources and ALMA SMGs in the training set used in this work. We also mark the flux densities that correspond to the 50\% completeness of S2COSMOS sources \citep{Simpson19}. The flux density of all ALMA archive SMGs (A$^{3}$COSMOS) are converted to $S_{870\,\mu m}$ if necessary using the flux ratios given in \cite{Fujimoto16}. We also present the flux distribution of ALMA SMGs in the UDS field (AS2UDS) for comparison. The flux density of our training sets covers that of single-dish S2COSMOS submillimeter sources.}
\label{f:flux_distribution.eps}
\end{figure}

\subsection{Machine-learning classification}
We apply the machine-learning classification in An18 to this work to identify the optical/NIR counterparts of S2COSMOS sources. In An18, we use two machine-learning algorithms, Support Vector Machine (SVM) \citep{Vapnik95} and XGBoost \citep{CG16}, to classify SMGs from field galaxies. The performance of these two algorithms is similar as shown in An18.  \cite{Liu19} tested a number of machine-learning algorithms and found most of them, including the deep learning, have very similar performance in differentiating SMGs from field galaxies. We should point out that the performances shown in \cite{Liu19} is slighter better than that in An18 and this work mainly because they used a smaller search radius ($r=7\arcsec$), which affect the final completeness of identification, since there are ALMA SMGs with offset $>7\arcsec$ from SCUBA-2 source as shown in Figure~\ref{f:radio.eps}.  Therefore, in this work, we prefer to retain  the two machine-learning algorithms we validated in An18 to identify multi-wavelength counterparts to SCUBA-2 sources in the COSMOS field. 

%
%
\startlongtable
\begin{deluxetable*}{lcccccc}
\tabletypesize{\scriptsize}
\tablecaption{Machine-learning performance based on different training sets\label{tab:table1}}
\tablehead{
\colhead{Training set/SVM} & \colhead{SMG\tablenotemark{a}} & \colhead{non-SMG\tablenotemark{a}} & \colhead{Recall (\%)} & \colhead{Precision (\%)} & \colhead{FPR (\%)} & \colhead{$F_{\rm 1} score (\%)$}
} 
\startdata
   AS2UDS & 255 & 1224 & $76\pm6$ & $76\pm7$  & $5.2\pm1.6$ & $76\pm6$ \\ 
   AS2COSMOS pilot & 100 & 850 & $81\pm7$ & $86\pm7$  & $1.4\pm0.9$ & $83\pm7$ \\ 
   A$^{3}$COSMOS &357  & 4186 & $81\pm4$ & $81\pm5$  & $1.6\pm0.5$ & $81\pm5$ \\ 
   AS2COSMOS$+$UDS & 355 & 2074 & $77\pm5$ & $80\pm5$  & $3.4\pm1.0$ & $78\pm5$ \\ 
   A$^{3}$+AS2COSMOS\tablenotemark{b} & 394 & 4509 & $83\pm4$ & $82\pm4$  & $2.0\pm0.6$ & $82\pm4$ \\
   \hline
Training set/XGB &   &  &  &  &  &  \\
 \hline
    AS2UDS & 364 & 1279 & $74\pm5$ & $81\pm4$ & $5.2\pm1.4$ & $77\pm5$ \\ 
   AS2COSMOS pilot & 126 & 928 & $78\pm9$ & $82\pm8$  & $2.2\pm1.0$ & $80\pm8$ \\ 
   A$^{3}$COSMOS & 445 & 4528  & $80\pm5$ & $81\pm4$  & $2.4\pm0.6$ & $81\pm5$ \\ 
   AS2COSMOS$+$UDS & 458 & 1999  & $74\pm4$ & $79\pm4$  & $4.4\pm1.1$ & $77\pm4$ \\ 
   A$^{3}$+AS2COSMOS\tablenotemark{b} & 490 & 4904 & $79\pm4$ & $82\pm4$  & $2.4\pm0.5$ & $81\pm4$ \\
 \enddata
\tablenotetext{a}{The number of SMG and non-SMGs in the training set;}
\tablenotetext{b}{Duplicates in these two training sets have been removed.}
\end{deluxetable*}

\subsubsection{Training set}
The effectiveness of machine-learning algorithms depends sensitively on the completeness and precision of the training set. On the one hand, a larger sample size provides better performance of the machine-learning classification. On the other hand, as we demonstrated in An18, differences of photometric systems between the training set and test samples will affect the performance of the machine-learning algorithms. Therefore, in this section, we compare the performance of both SVM and XGBoost classifiers by using training sets based on three different ALMA surveys. We described two of them, AS2COSMOS pilot  and A$^{3}$COSMOS, in Section \ref{sec:obs_ts}. The other ALMA survey we use is the 870\,\mm\ ALMA survey of 716 SCUBA-2 sources in the UDS field \citep[AS2UDS;][]{Stach19}, which was used to build the training set in An18.

We show the flux density distributions of ALMA SMGs in these three ALMA surveys in Figure~\ref{f:flux_distribution.eps}. The flux densities of A$^{3}$COSMOS SMGs are converted to $S_{\rm 870\,\mu m}$ by adopting the ratios estimated by \cite{Fujimoto16}.

Following An18, a non-SMG is defined as any optical/NIR source that lies within the ALMA primary beams but does not have a secure detection from ALMA (e.g., does not have $\ge 4.8\,\sigma$ detection in the AS2COSMOS pilot). In An18, we  identified that the photometric redshift ($z_{\rm phot}$), absolute $H$-band magnitude ($M_{H}$) and NIR colors are the most efficient properties for differentiating SMGs from field galaxies. The COSMOS field has deep IRAC 3.6\,\mm\ and 4.5\,\mm\ data along with the $z^{++}$-band data, which is used to detect sources outside the regions of UltraVISTA-DR2 by \cite{Laigle16}. Hence the NIR colors we choose in this work are ($z-$[3.6]) and ([3.6]-[4.5]). While the SVM classifier requires detection in all the selected properties, the XGBoost classification can be performed with missing features. Therefore, when constructing the training sets for XGBoost, we do not require secure detection in $z^{++}$-band. In addition, we include ($J-K_{\rm s}$) and ($K_{\rm s}-$ [3.6]) colors if the source has secure detections in the corresponding band(s).

We list the number of SMGs and non-SMGs with the secure measurements of selected features in each training set in Table~\ref{tab:table1}. Without the limitation of detection in $z^{++}$-band, the sample sizes of XGBoost increase $\sim$10--15\% compared to that of SVM. We also combine the training sets based on AS2UDS, AS2COSMOS and A$^{3}$COSMOS surveys to enlarge the training set. The duplicates between AS2COSMOS pilot and A$^{3}$COSMOS have been removed in the combined training set to guarantee a uniform weight for all sources when training the machine-learning classifiers. 

\begin{figure*}[!t]
\plottwo{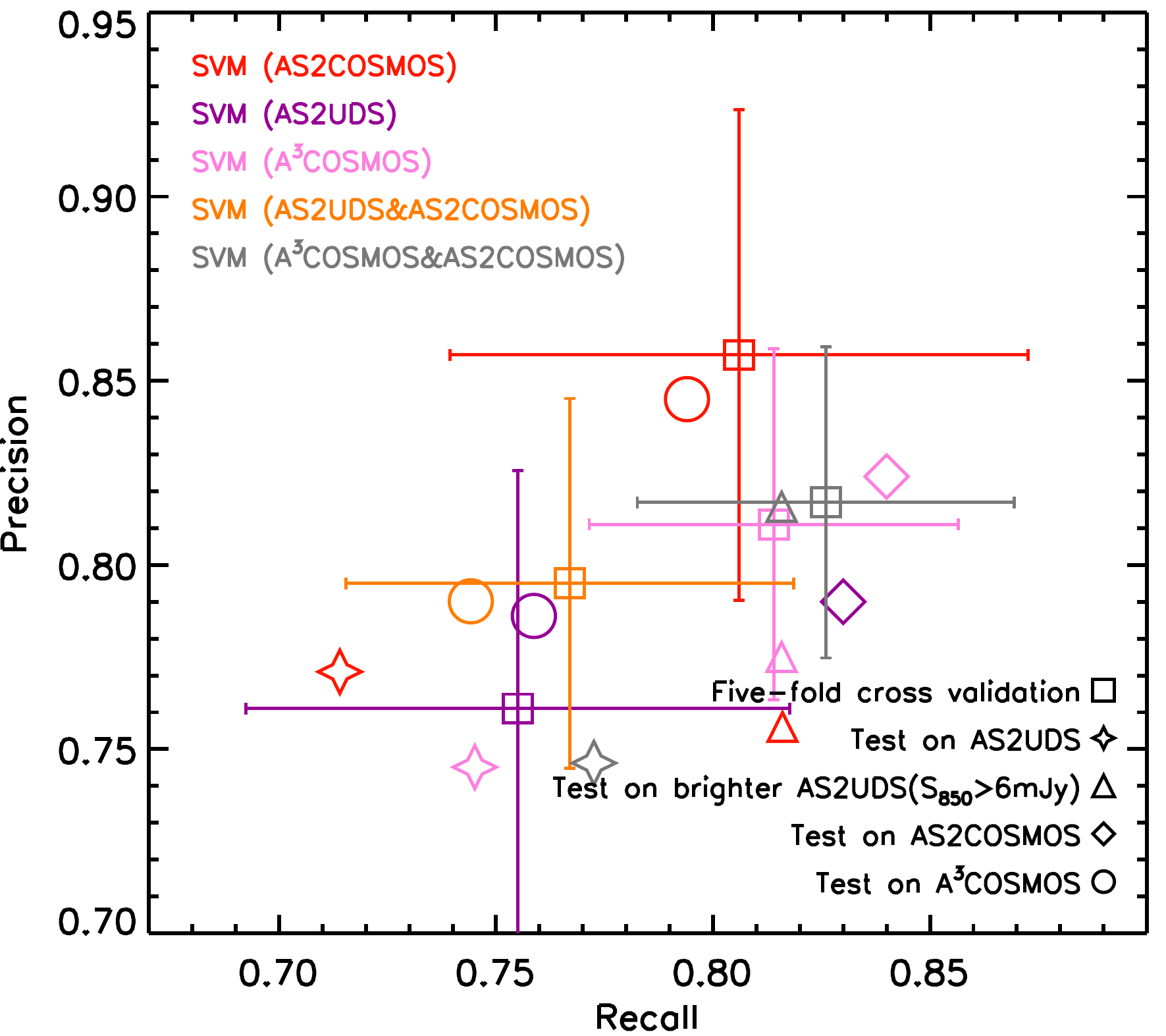}{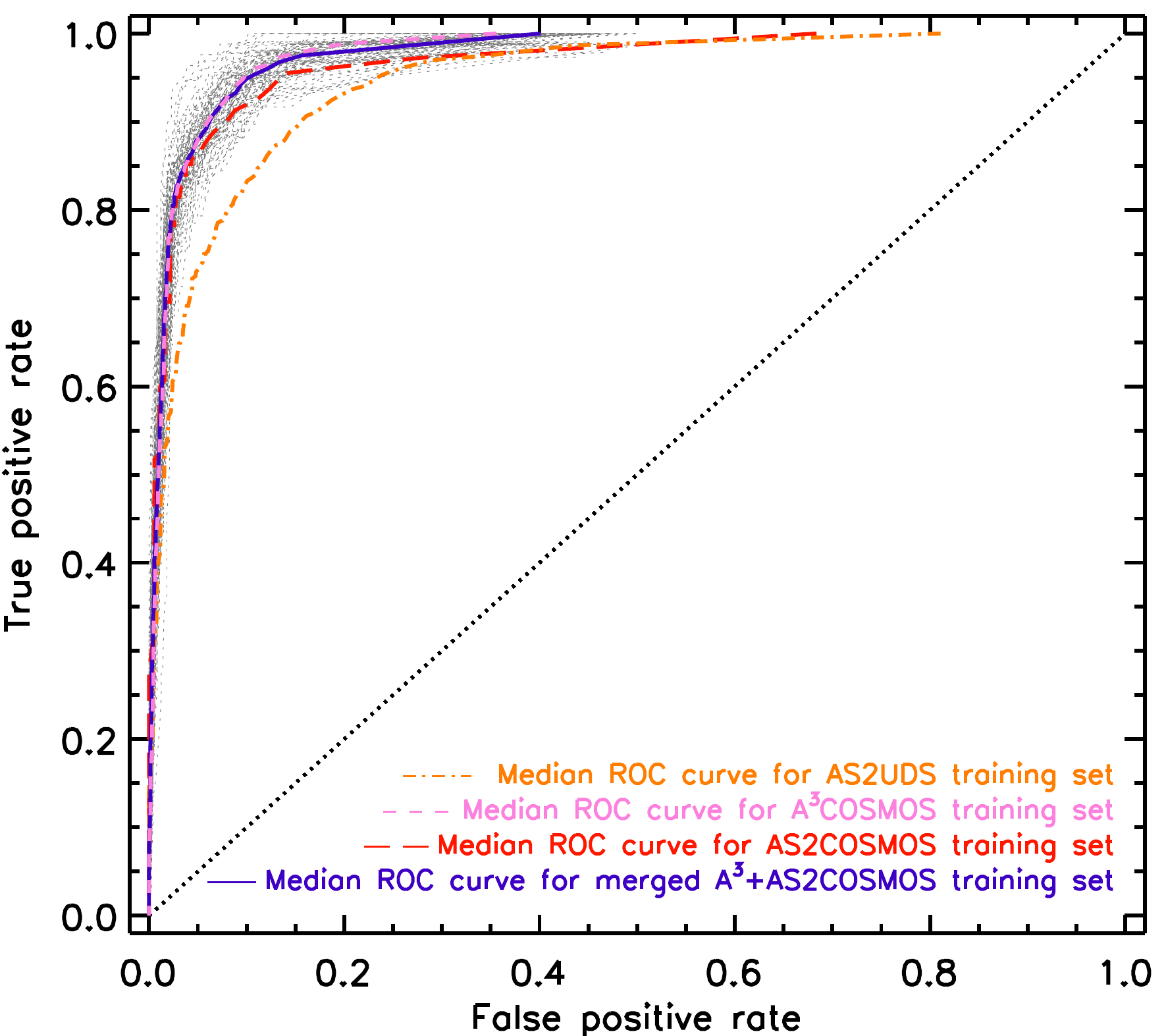}
\caption{{\it Left: }Comparison of the SVM classification performances based on different training sets. We use different colors to show the results based on different training sets and different symbols represent the results of different tests. For each training set, we use five-fold cross-validation to optimize the parameters of the SVM classifier and show the optimized results as squares in the plot, where the errors are standard deviation of precision and recovery rate based on 100 bootstrap-simulations. Then we test the performance of the SVM classifier based on a different training set by using the other training sets as the test sample.  Through these comparisons, we find the SVM based on the combined A$^{3}$COSMOS and AS2COSMOS pilot training sets has the best performance in differentiating SMGs from field galaxies. {\it Right:} Comparison of the Receiver Operating Characteristic (ROC) curves for optimized XGBoost classifier trained on different training sets. The XGBoost classifier is optimized through the five-fold cross-validation. We repeat the five-fold cross-validation 100 times, calculate the median true and false positive rates and use them to represent the curve. As shown in the plot, the ROC curve based on merged A$^{3}$COSMOS and AS2COSMOS pilot training set has the maximal area-under-the-curve (AUC). Therefore, this merged sample is adopted as the training set in this work. The background grey dotted lines represent the individual ROC curves of five-fold cross-validation based on this merged training set.\label{f:svm.eps}}
\end{figure*}

The parameters of both SVM and XGBoost classifiers are optimized by five-fold cross-validation \citep{Kohavi95}, which means we first divide the training set into five subsets and train the machine-learning classifier on four folds and validate on the remaining one. We repeat the five-fold cross-validation 100 times to estimate the scatter in evaluation metrics.
The evaluation metrics we use in this work are {\it Recall}, {\it Precision}, False Positive Rate (FPR) and the $F_{\rm1}$ score. We have defined {\it Precision} and {\it Recall} in Section~\ref{s:identification_radio}. FPR is defined as the number of sources that are incorrectly classified as SMGs over the total number of non-SMGs in the data set. We also use the $F_{\rm1}$ score, which is defined as 2$\times$(Precision$\times$Recall)/(Precision$+$Recall), as one of the metric to evaluate the performance of machine-learning classifier trained by different training sets. We show the {\it Recall} and {\it Precision} of five-fold cross-validation with the optimized SVM classifiers trained on different training sets in the left panel of Figure~\ref{f:svm.eps} and list the values of all evaluation metrics in Table~\ref{tab:table1}.

The SVM classifier trained by the sample based on AS2UDS has the lowest $F_{\rm1}$ score 76\%. Possible reasons for this are that the training set based on AS2UDS comprises the $K$-band-selected sources, and that the 3.6\,\mm\ and 4.5\,\mm\ observations in the UDS field are shallower than those in the COSMOS field. Therefore, the selected input features based on the multi-wavelength photometry in the COSMOS field are not the best properties for the photometric systems in the UDS field to train the machine-learning classifier. For instance, the $F_{\rm1}$ score of five-fold cross-validation of the AS2UDS training set with the different input features in An18 is $\sim$80\%. The SVM classifier trained by A$^{3}$COSMOS training sets has a performance with $F_{\rm1}=81$\% as shown in Table~\ref{tab:table1}. The performance of SVM is slightly increased to $F_{\rm1}=82$\% if we combine the training sets of AS2COSMOS pilot and A$^{3}$COSMOS (A$^{3}+$AS2COSMOS). Although the SVM classifier trained by AS2COSMOS pilot shows the best performance with $F_{\rm1}=83$\%, the scatter is larger because of the small sample size. In addition, the AS2COSMOS pilot corresponds to the brighter single-dish submillimeter sources ($S_{850\,\mu m}>$ 6\,mJy). Therefore, we also evaluate the SVM classifiers that are trained on different training sets by using the other training set as the test sample and show the results in the left panel of Figure~\ref{f:svm.eps}. We find that if we test the machine-learning method on the test sample that corresponds to the brighter single-dish submillimeter sources, i.e., AS2COSMOS pilot and brighter AS2UDS, the results will be overestimated. As shown in Figure~\ref{f:svm.eps}, if we use the training set that are constructed by the brighter SCUBA-2 sources, then we will fail to recover $\sim$5\% SMGs with moderate/fainter submillimeter emission. We also notice that if we use the training set from UDS field to recover SMGs in the COSMOS field and vice versa, the differences between observation depth, source-selection, and the estimates of the photometric redshift will cause a $\sim$4--6\% decrease in the success of machine-learning classification. Overall, the SVM classifier trained on A$^{3}+$AS2COSMOS has better performance in all cases as shown in Figure~\ref{f:svm.eps}.

%
%
\begin{figure*}
\centering
\includegraphics[width=0.98\textwidth]{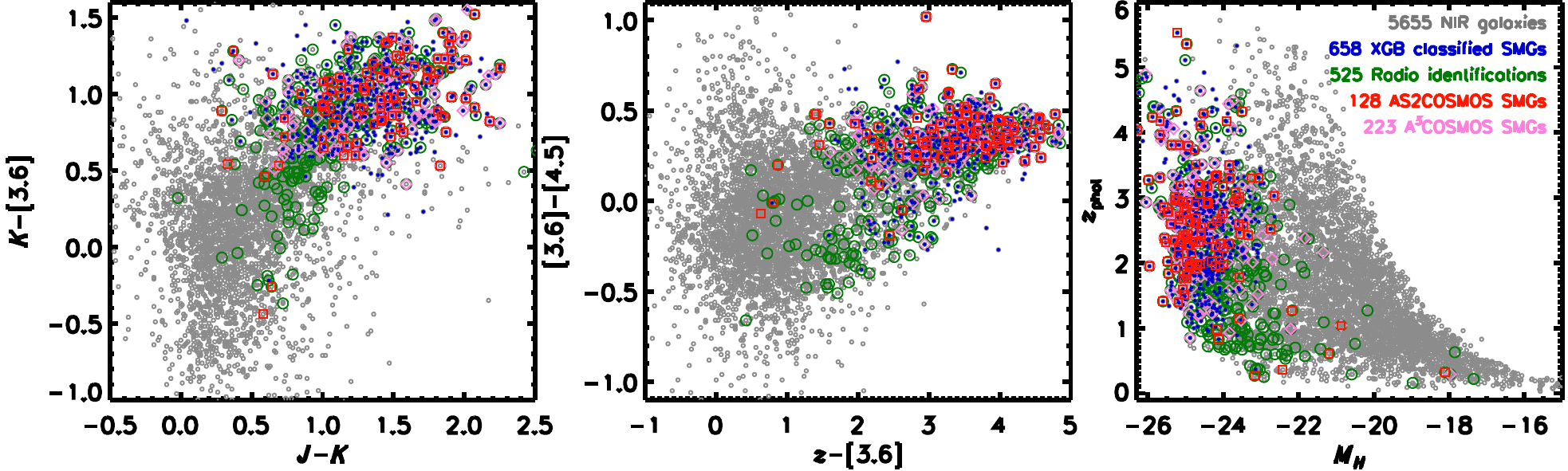}
\caption{Results of applying the XGBoost classifier to identify counterparts to SCUBA-2-detected submillimeter sources in NIR observations of the COSMOS field. The grey circles are 5,655 NIR-selected galaxies, which are located within 1,066 SCUBA-2 error circles and have secure detection in 3.6\,\mm\ and 4.5\,\mm-band and sufficient photometry for estimating the redshift and absolute $H$-band magnitude. The blue points represent the 658 counterparts of SCUBA-2 submillimeter sources classified by the machine-learning. The red squares show the 128 ALMA SMGs from our pilot ALMA follow-up observations of brighter ($S_{850\,\mu m} >6$\,mJy) S2COSMOS submillimeter sources. XGBoost recalls 88\% ALMA SMGs which corresponded to brighter single-dish submillimeter sources. There are also 223 ALMA SMGs from the ALMA archive located within the SCUBA-2 error circles that meet the requirement of machine-learning classification. XGBoost recovers 90\% of these ALMA SMGs. However, we note that all of these ALMA SMGs (AS2COSMOS pilot and A$^{3}$COSMOS) are part of the training set. Therefore, the {\it Recall} is higher than that of five-fold cross validation of the training set, which is $\sim$80\%. We also mark the 525 NIR-selected galaxies with $>5\,\sigma$ 3\,GHz radio detection and $p\le0.065$, i.e., radio identified counterparts to SCUBA-2 sources. There are 368 NIR galaxies classified as the SMGs by both radio and machine-learning.  Combining with the radio identification, in total, we identify 815 counterparts of SMGs from 5,655 NIR-selected galaxies in the COSMOS field. According to the ``self-test'' of the AS2UDS sample and the deeper radio observation in COSMOS, we expect that our radio combined machine-learning method recovers $\ge$85\% ALMA SMGs from these 5,655 NIR galaxies.}
\label{f:machine_learning.eps}
\end{figure*}

We do the same evaluation for the XGBoost classifier. We show the Receiver Operating Characteristic (ROC) curves from the optimized XGBoost classifier trained on different training sets in the right panel of Figure~\ref{f:svm.eps}. The ROC curves are constructed by comparing the {\it Recall} against the FPR, as the probability threshold is varied \citep{Fawcett04}. We then use the area under the curve (AUC) of a ROC curve \citep{Fawcett04} and the evaluation metrics shown in Table~\ref{tab:table1} to evaluate the performance of XGBoost classifiers. The classifier trained on the A$^{3}+$AS2COSMOS training set has a better performance with the maximal AUC (Figure~\ref{f:svm.eps}) and $F_{\rm1}=81$\%. Therefore, we choose the training set based on the A$^{3}+$AS2COSMOS surveys to train the machine-learning classifier in this work.

For the two machine-learning algorithms, SVM has a slightly better performance compare to XGBoost as shown in Table~\ref{tab:table1}. However, the SVM can not deal with missing features unless they are artificially filled by imputated data. In this work, the primary missing features are NIR colors of sources. Unfortunately, the cause of the lack of measurement of NIR colors are various. They could be due to dust reddening, geometry, star-formation history, redshift and so on. For classifying SMGs and non-SMGs, the statistical imputation techniques do not improve the performance of machine-learning classifiers and even make the results worse in some cases, as demonstrated in \cite{Liu19}. In contrast, the XGBoost classifier can perform classification with missing features by classifying the instance into the optimal default direction that is learned from the data \citep{CG16}. If we adopt the XGBoost classifier, the sample size of both training set and test sample increase without the limitation of requiring secure detection in $z^{++}$-band, as shown in Table~\ref{tab:table1}.  And these two classifiers have a very similar precision while the recall of SVM is $\sim$2--4\% higher than that of XGBoost. However, the sample size of XGBoost is $\sim$10\% larger than that of SVM. Therefore, the final completeness of the machine-learning method is still higher if we adopt the XGBoost classifier. We therefore prefer to use the XGBoost classifier in this work to increase the sample size of both training set and test sample and hence the completeness of identified counterparts of S2COSMOS submillimeter sources. 

%
%
\startlongtable
\begin{deluxetable*}{lccccccccr}
\tabletypesize{\scriptsize}
\tablecaption{Radio$+$machine-learning identified counterparts to S2COSMOS sources \label{tab:all_id}}
\tablehead{
\colhead{ID$_{\rm S2}$} & \colhead{RA$_{\rm S2}$} & \colhead{DEC$_{\rm S2}$} & \colhead{RA$_{\rm NIR}$ } & \colhead{DEC$_{\rm NIR}$} & \colhead{Flag$_{\rm xgb}$\tablenotemark{a}} & \colhead{RA$_{\rm radio}$} & \colhead{DEC$_{\rm radio}$} & \colhead{Flag$_{\rm radio}$\tablenotemark{b}} & \colhead{$z_{\rm phot}$}
} 

\startdata
S2COS850.0001   &150.033530   &2.436552  &... &...  &...  &150.033508   &2.436735     &1     &... \\
S2COS850.0002   &150.064676   &2.263777  &150.065170   &2.263673     &1  &150.065063   &2.263606     &1       &3.31 \\
S2COS850.0003   &150.238150   &2.337106  &150.238708   &2.336827     &1  &150.238647   &2.336836     &1       &2.29 \\
S2COS850.0003   &150.238150   &2.337106  &150.239273   &2.336381     &1  &...  &...  &...       &2.36 \\
S2COS850.0006   &149.989043   &2.458214  &149.988724   &2.458332     &0  &149.988693   &2.458483     &1       &3.30 \\
... & ... & ... & ... & ... & ... & ... & ... & ... & ...  \\
 \enddata
 \tablenotetext{a}{Flag$_{\rm xgb}=1$: XGBoost classified SMGs; 0: classified non-SMGs; ...: not qualified for machine-learning classification;}
 \tablenotetext{b}{Flag$_{\rm radio}=1$: radio identified SMGs; ...: do not have radio detection;}
 \tablecomments{This table is available in its entirety in machine-readable form.}
\end{deluxetable*}

\subsubsection{Machine-learning results}
Having selected a training set and machine-learning algorithm, we then use the optimized XGBoost classifier to identify optical/NIR counterparts to S2COSMOS submillimeter sources. For the 1,147 S2COSMOS sources, 1066 of them are within optical/NIR coverages. Within their error circles, there are 5,655 NIR-selected sources that have secure detection in IRAC 3.6\,\mm\ and 4.5\,\mm-band and also have estimated photometric redshift and absolute $H$-band magnitude in the COSMOS2015 catalog. These are used to construct the test sample for classifying counterparts to S2COSMOS sources. We show the results of classification in Figure~\ref{f:machine_learning.eps}. The XGBoost classifier identifies 658 counterparts of S2COSMOS sources from the 5,655 NIR-selected sources. Among them, 368 also have radio identified counterparts. According to the self-test of the AS2UDS data set in An18, the {\it Precision} of radio \& machine-learning classification could reach 90\%. Combining with radio identification, in total, we identify 815 optical/NIR/radio counterparts to S2COSMOS submillimeter sources from the 5,655 NIR-selected sources. The expected {\it Recall} of the radio combined machine-learning method in this work is $\ge$85\%, according to the self-test in An18 and the relatively deeper radio observation in the COSMOS field.

In Section~\ref{s:identification_radio}, we identified 932 radio counterparts to S2COSMOS sources by using the $p$-value to calculate the probabilistic association of radio sources to S2COSMOS sources. Of these 368 are also identified by machine-learning, while radio identification alone find 564 counterparts to S2COSMOS sources. Therefore, in total, our radio combined machine-learning method identifies 1,222 optical/NIR/radio counterparts of single-dish detected submillimeter sources in the COSMOS field, which are listed in Table~\ref{tab:all_id}. 

%
%
\begin{figure}
\centering
\includegraphics[width=0.46\textwidth]{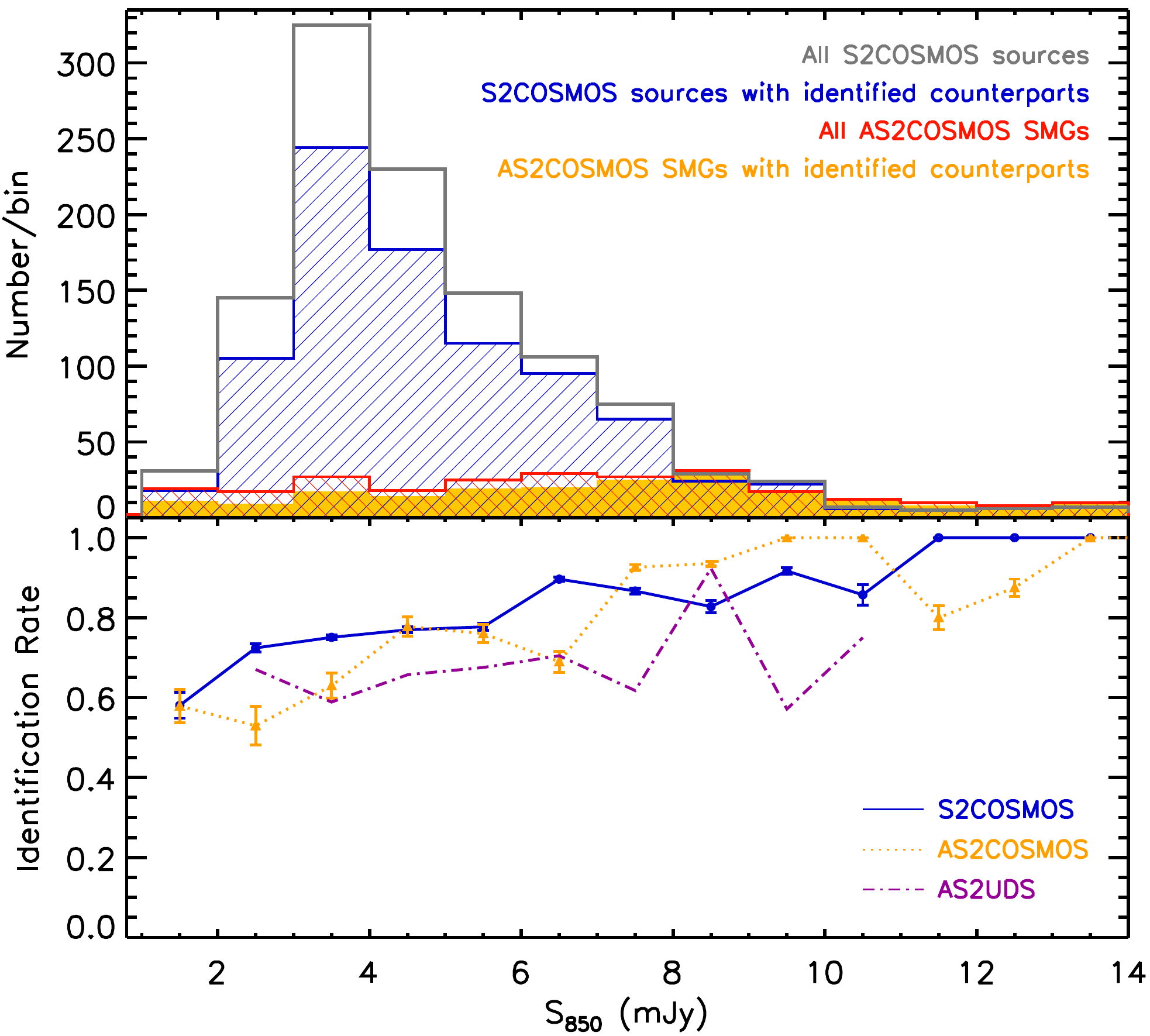}
\caption{The identification rate as a function of submillimeter flux density. {\it Top:} The flux density distributions of all SCUBA-2 sources, SCUBA-2 sources having at least one identified counterpart, and AS2COSMOS pilot SMGs having identified radio or NIR counterpart. {\it Bottom:} The fraction of SCUBA-2 sources with at least one identified counterpart and the fraction of ALMA SMGs with identified counterpart  as a function of $S_{\rm 850\,\mu m}$. It can be seen that the identification rate increases for both brighter single-dish-detected submillimeter sources and ALMA-detected SMGs. We also show the completeness of the radio+machine-learning method of AS2UDS sample as a function of $S_{\rm 850\,\mu m}$ for comparison. We expect a slightly higher completeness of S2COMSOS SMGs because of deeper radio data compared to that in the UDS field.}
\label{f:identification.eps}
\end{figure}

\subsubsection{Identification rate}
The completeness of the radio$+$machine-learning method is expected to be 64\% based on the test of AS2UDS in An18. This fraction should be higher for brighter submillimeter sources because of the increased identification rate as shown in Figure~\ref{f:identification.eps}. The identification rate is defined as the number of single-dish submillimeter sources having at least one identified counterparts versus the total number of single-dish submillimeter sources in the survey. For the 1,145 S2COSMOS sources with radio or NIR coverage, our method identified multi-wavelength counterparts for 897 of them. Therefore, the average identification rate is 78\%. From Figure~\ref{f:identification.eps}, we can see that the identification rate increases with the flux density of S2COSMOS sources. We also show the identification rate of AS2COSMOS pilot SMGs as a function of $S_{\rm 870\,\mu m}$ in Figure~\ref{f:identification.eps}. The identification rate of ALMA SMGs also increases with submillimeter flux density.  The identification rate of AS2UDS SMGs in An18 is shown for comparison. On average, the identification rate of S2COSMOS sources is higher than that of AS2UDS SMGs, which is most likely caused by the false positive identifications of the radio and machine-learning method. The expected identification rate of SMGs within the SCUBA-2 error circles in the COSMOS field is slighter higher than that in the UDS field because of the slightly deeper radio data.

\section{Results and Discussion}\label{s:main_results}
Having identified multi-wavelength counterparts of S2COSMOS sources, we now analyze the physical properties of these single-dish-detected submillimeter sources. 

\subsection{Redshift distribution}\label{s:Redshift_distribution}

\begin{figure}
\centering
\includegraphics[width=0.46\textwidth]{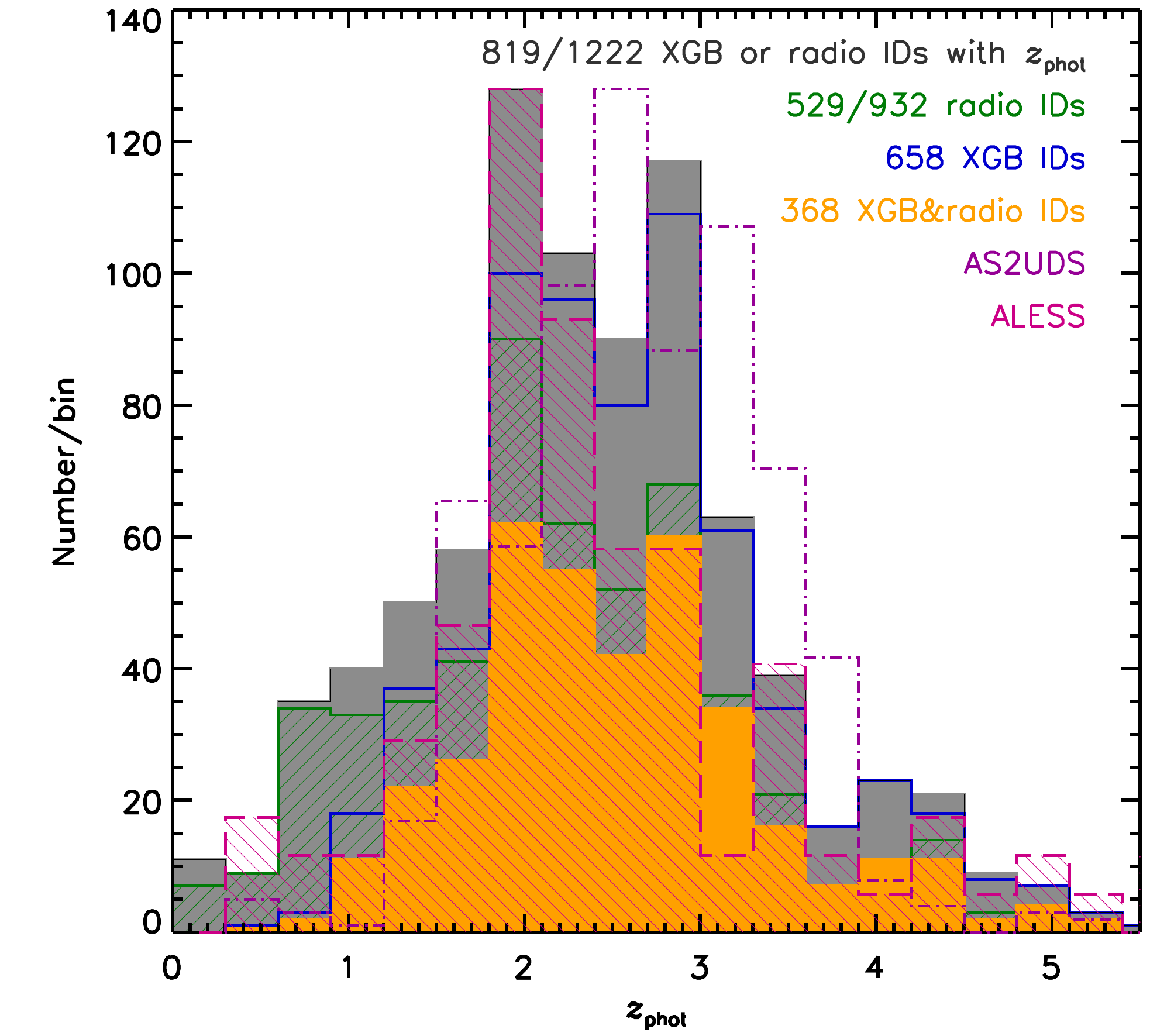}
\caption{The redshift distributions of radio+machine-learning identified counterparts to SCUBA-2 detected submillimeter sources in the COSMOS field. We plot the distribution of all 819/1,122 XGBoost or radio identifications with photometric redshift ($z_{\rm phot}$). We also separately present the redshift distribution of radio and machine-learning identified counterparts to submillimeter sources. The distribution of galaxies that are identified as counterparts of SMGs by both radio and machine-learning is shown as the orange area. For comparison, we also show the redshift distribution of ALMA SMGs in the UDS field (AS2UDS, U. Dudzevi{\^c}i{\=u}t{\.e} et al.\ 2019, in preparation) and the ECDFS field \citep[ALESS,][]{Simpson14}. The distributions of these two ALMA samples are scaled to compare with our results. In general, our radio$+$machine-learning identified counterparts of SMGs have a similar redshift distribution to ALMA-detected SMGs. }
\label{f:redshift_distribution.eps}
\end{figure}

For the 1,222 radio$+$machine-learning identified counterparts, 819 have estimated photometric redshifts from the COSMOS2015 catalog with $z_{\rm median}=2.3\pm0.1$. 
We show the redshift distribution of our identified optical/NIR/radio counterparts of S2COSMOS sources in Figure~\ref{f:redshift_distribution.eps}.

By comparing the redshift distribution of machine-learning or radio identified counterparts with that of ALMA SMGs in the UDS (AS2UDS) and the ECDFS (ALESS) fields, we find that the machine-learning fails to recover some low-redshift counterparts while the radio identification includes some low-redshift contamination. For the machine-learning classification, photometric redshift is one of the input features in the analyses. Therefore, the trained classifier tends to recover SMGs at high redshift. Radio observations do not benefit from a negative $k$-correction, hence, they will miss SMGs at high redshift and preferentially include some contaminations at low redshift. Nevertheless, the overall redshift distribution of radio$+$machine-learning identified counterparts of S2COSMOS sources is broadly consistent with that of ALMA SMGs \citep[][U. Dudzevi{\^c}i{\=u}t{\.e} et al.\ 2019, in preparation]{Simpson14} as shown in Figure~\ref{f:redshift_distribution.eps}. 

%
%

%
%
\begin{figure}[!t]
\centering
\includegraphics[width=0.46\textwidth]{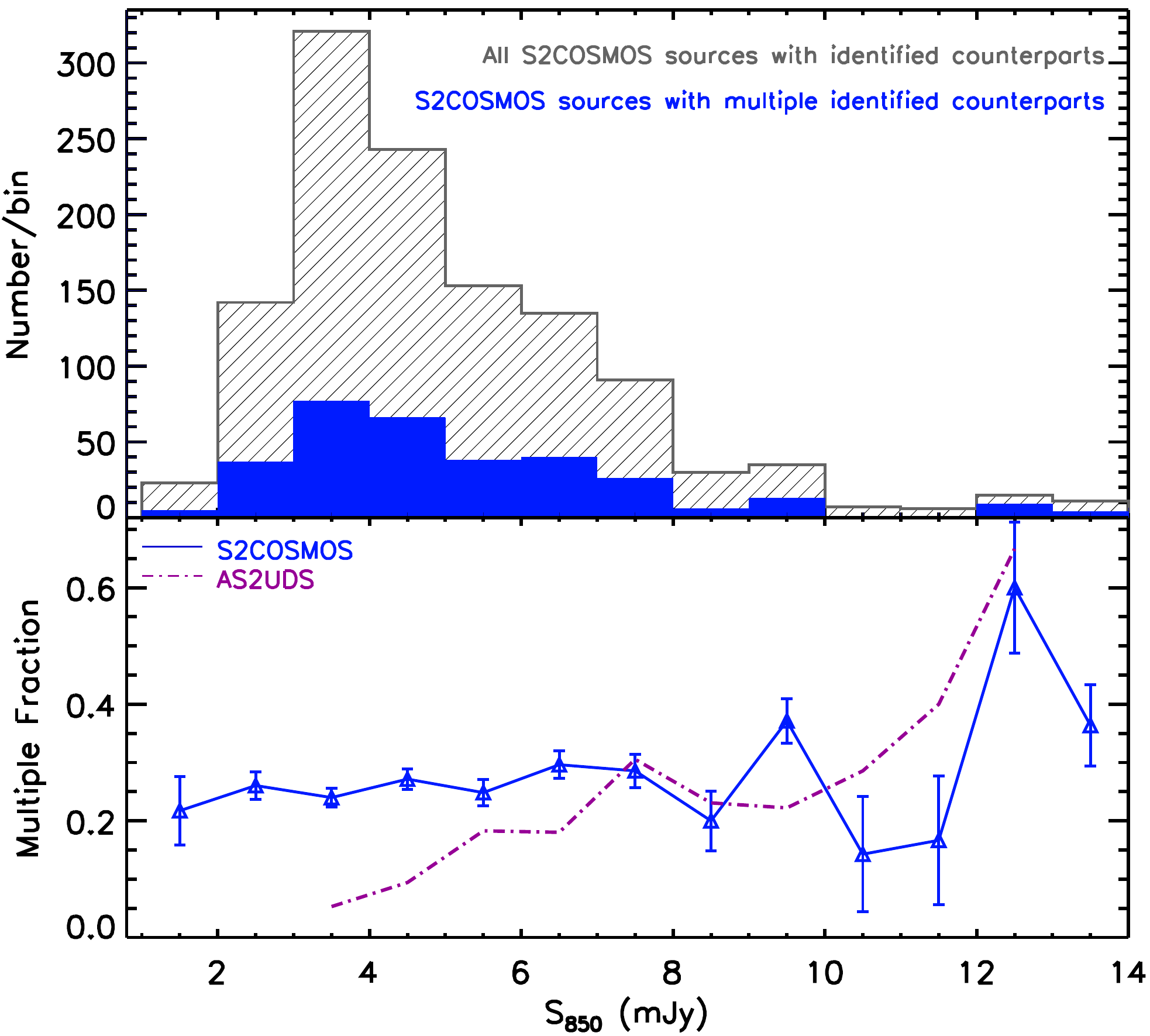}
\caption{{\it Top:} The flux density distribution of all SCUBA-2 sources that have radio or machine-learning identified counterparts in the COSMOS field and the distribution of SCUBA-2 sources with multiple identified counterparts. {\it Bottom:} The fraction of  SCUBA-2 sources with multiple identified counterparts as a function of $S_{850\,\mu m}$. On average, the multiple fraction is ($26\pm5$)\% and this fraction increases for brighter single-dish-detected sources. We also show the multiple fraction of the AS2UDS sample for comparison. The higher multiple fraction at the faint-end of this work might be caused by the false positive detections of our radio$+$machine-learning method. However, our stacking analyses shown in \cite{An18} confirm that the machine-learning method can identify the faint/diffuse submillimeter emissions even below the ALMA detection limit. The accuracy of multiplicity of S2COSMOS sources is affected by both incompleteness and false-identifications of the radio$+$machine-learning method.}
\label{f:multiple_rate.eps}
\end{figure}

%
%
\begin{figure}
\centering
\includegraphics[width=0.46\textwidth]{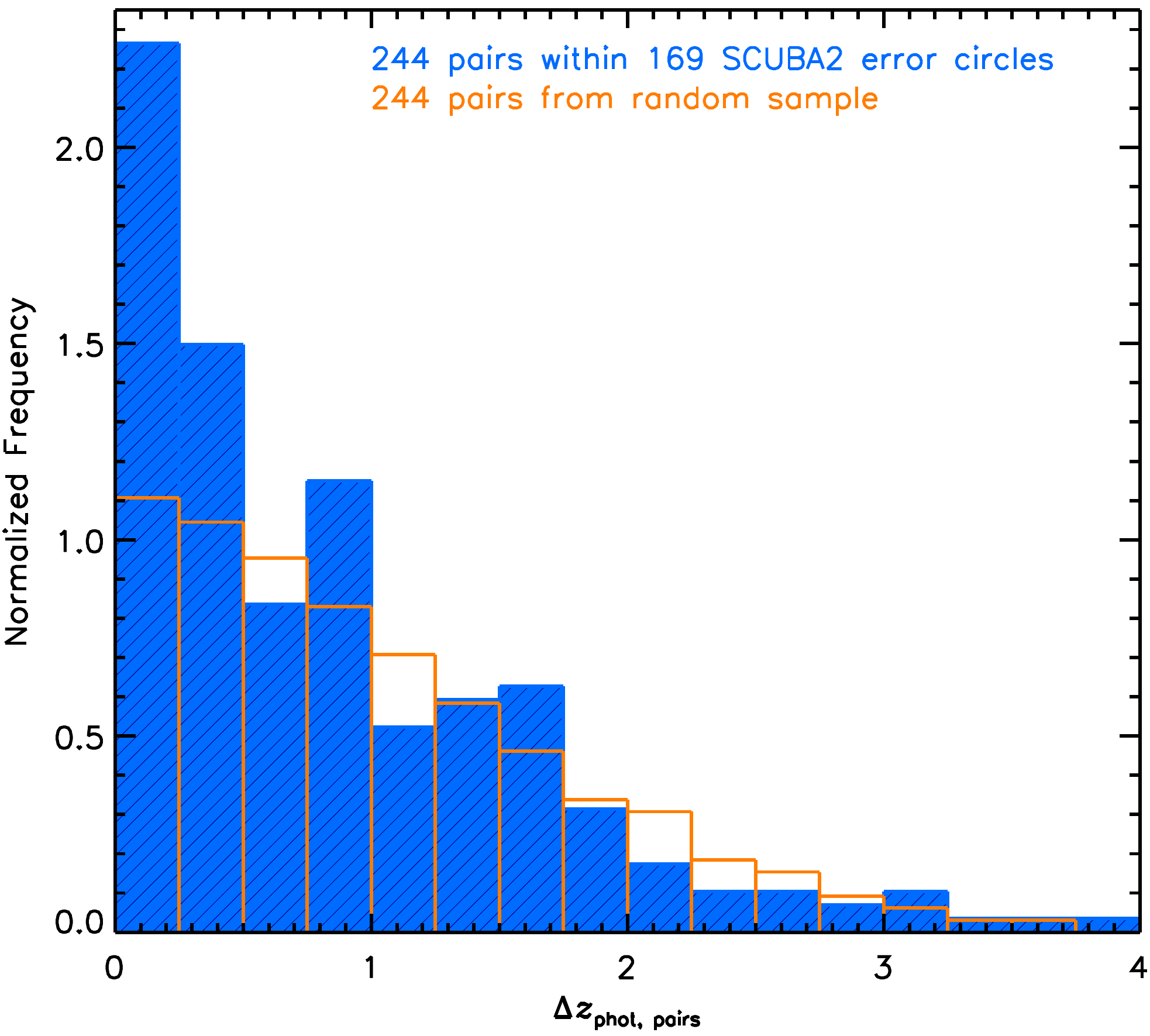}
\caption{The normalized distribution of $\Delta z$ for the pairs of radio$+$machine-learning identified counterparts within the same SCUBA-2 error circle, compared to the pairs randomly selected from the distribution of all isolated counterparts with photometric redshift. The distribution is normalized by assuming that each SCUBA-2 source only has a single counterpart. The strong peak at $\Delta z<0.25$ for the identified counterparts pairs compared with a random sample indicates that a moderate fraction of multiple counterparts in the same SCUBA-2 error circles arises from physically associated galaxies, rather than chance line-of-sight projections.}
\label{f:multiplicity.eps}
\end{figure}

\subsection{Multiplicity}
Recent studies using interferometric observations in submillimeter/milimeter suggest $\gs$20\% of single-dish-detected submillimeter sources actually correspond to blends of multiple SMGs \citep[e.g.,][]{Wang11, Karim13, Simpson15a,Simpson15b, Stach18}. In this work, multiplicity is defined as a single-dish-detected submillimeter sources having more than one identified counterparts from our radio$+$machine-learning method. We show the multiple fraction as a function of flux density of single-dish-detected submillimeter sources in Figure~\ref{f:multiple_rate.eps}. The average fraction of S2COSMOS sources with multiple identified counterparts is ($26\pm5$)\%. This fraction is fairly constant, and only slightly increases for brighter single-dish submillimeter sources as shown in Figure~\ref{f:multiple_rate.eps}.

\cite{Stach18} in their ALMA AS2UDS survey studied the multiplicity of S2CLS-UDS single-dish submillimeter sources and found a multiplicity of ($28\pm2$)\% for submillimeter sources with $S_{\rm 850\,\mu m}\ge 5$\,mJy and of ($44^{+16}_{-14}$)\% at $S_{\rm 850\,\mu m}\ge 9$\,mJy. We plot the multiplicity based on AS2UDS in Figure~\ref{f:multiple_rate.eps} for comparison. The multiple fraction of S2COSMOS sources increases with the flux density of the SCUBA-2 source and is higher than that of AS2UDS at the faint-end, which most likely is a result of incompleteness in identifying faint companions in those ALMA maps. In An18, we stacked the submillimeter emission in ALMA maps at the position of the machine-learning classified but individually ALMA-undetected NIR-selected galaxies and found that the machine-learning can recover the faint submillimeter emissions even if they are below the detection threshold of ALMA observations ($S_{\rm 870\,\mu m}\sim 1$--2\,mJy). This is further confirmed by deeper ALMA observations in Cycle 5 for the ten brightest SCUBA-2 sources in AS2UDS, which did not have any secure ALMA detections in Cycles 3 and 4 \citep{Stach19}. These deeper observations recovered counterparts to nine of the ten target fields. This may explain the decline in  multiplicity at the faint-end in AS2UDS shown in Figure~\ref{f:multiple_rate.eps}. However, the false positive detection of the radio$+$machine-learning method may also cause a modest increase of multiplicity of S2COSMOS sources, although this isn't expected to depend on $S_{\rm 850\,\mu m}$.

As described in Section \ref{s:Redshift_distribution}, 819/1222 radio$+$machine-learning identified counterparts have estimated photometric redshift from the COSMOS2015 catalog \citep{Laigle16}. Among their corresponding single-dish SCUBA-2 sources, our study of multiplicity shows that 133 (15\%) of them have two, 35 (4\%) have three and one has four radio$+$machine-learning identified counterparts. For the SCUBA-2 sources that have multiple SMGs, \cite{Simpson15a} and \cite{Stach18} suggested that $\sim30\%$ of them arise from physically associated galaxies based on their photometric redshift ($\Delta z_{\rm phot}$), while \cite{Wardlow18}, using an ALMA CO survey of six single-dish submillimeter sources, found that (36$\pm$18)\% multiple SMG components in blended single-dish submillimeter sources are closely physically associated. 

In this work, we show the normalized distribution of $\Delta z_{\rm phot}$ for the 244 pairs that are identified as counterparts to the same S2COSMOS sources in Figure~\ref{f:multiplicity.eps}. We find that 65 of 244 pairs (27\%) have $\Delta z_{\rm phot}<0.25$. The choice of $\Delta z_{\rm phot}<0.25$ is to compare with the results of ALMA SMGs in \cite{Stach18}. We also note that the choice of $\Delta z_{\rm phot}<0.25$ corresponds to $\sim2.5\,\sigma$ of the uncertainty of the photometric redshift of radio$+$machine-learning identified counterparts of SMGs as shown in Section \ref{s:Redshift_distribution}. To test the significance of this result, we randomly select 244 pairs from the 1,222 identified counterparts of S2COSMOS sources. We repeat this random selection 100 times and show the median value of the distribution of $\Delta z$ for these random pairs in Figure~\ref{f:multiplicity.eps} for comparison. On average, 15\% of the random pairs have $\Delta z_{\rm phot}<0.25$, which is half of that for identified counterpart pairs. This suggests that a moderate fraction ($\sim27\%$) of multiple counterparts to the same SCUBA-2 sources arise from physically associated galaxies, rather than line-of-sight projections by chance, although our result is affected by the incompleteness and false positive identification of the radio$+$machine-learning method.

\subsection{AGN fraction}
Both theoretical simulations and observational studies suggest a link between the growth of galaxies and their supermassive black holes \citep[SMBHs, $M_{\rm BH}=10^{6}$--$10^{9}\Msun$, e.g.,][]{Hopkins08, Ishibashi16}. Under this paradigm, starburst-dominated galaxies and the active galactic nuclei (AGN) dominated QSOs are essentially the same systems observed at different 
evolutionary stages \citep[e.g.,][]{Sanders88,Perna18}. Therefore, surveying the AGN activity in the SMG population provides insights into not only SMBH growth, but also potentially the evolutionary cycle of massive galaxies. 

We start by looking at the AGN population in radio-identified SMGs. We take   advantage of the fact that \cite{Smolcic17b} has identified AGNs from the VLA-COSMOS 3\,GHz radio data using the X-ray, mid-infrared (MIR) color-color and multi-wavelength SEDs selection methods. For the 932 radio-identified counterparts of S2COSMOS sources, 80 have X-ray detections in the $Chandra$ COSMOS-Legacy Survey \citep{Civano16, Marchesi16}. Among these, 74 are classified as X-ray AGNs according to their X-ray luminosity \citep[$L_{\rm [0.5-8]\,Kev}>10^{42}$\,erg\,s$^{-1}$, although we note that this is conservative for strongly star-forming galaxies,][]{Smolcic17b}. Since the X-ray selection of AGN is progressively missing faint AGN at high redshift because of the limitation of survey depth, \cite{Smolcic17b} also adopt the MIR color-color selection method of \cite{Donley12} and SED-selection to complement the X-ray selection criterion. Of the radio counterparts of S2COSMOS sources with $z_{\rm phot}\ls3.0$, 125 meet the MIR AGN criterion, although 41 of them are also X-ray AGNs. \cite{Smolcic17b} also classified radio sources that show AGN signatures in their optical to millimeter SED as SED-selected AGNs. Among these, 163 are counterparts of S2COSMOS sources. In total, 225/932 ($24\pm4$)\% radio-identified counterparts of S2COSMOS sources are classified as AGNs in \cite{Smolcic17b}. 

For the remaining 290 optical/NIR counterparts of S2COSMOS sources, which lack radio counterparts and are classified by machine-learning, none of these have X-ray detections in {\it XMM}-COSMOS  \citep{Hasinger07,Cappelluti07,Brusa10} or $Chandra$ surveys \citep{Civano16, Marchesi16}. Therefore, the fraction of X-ray detected AGN in our radio$+$machine-learning identified counterparts of S2COSMOS sources is ($6\pm2$)\%. We also adopt the MIR color-color selection criteria in \cite{Donley12} for the 170/290 optical/NIR counterparts of SCUBA-2 sources, which have $z_{\rm phot}\le3.0$ and $>3\,\sigma$ detections in all four IRAC bands \citep[see][]{Stach19}. Three of these meet the selection criteria of AGN from \cite{Donley12}. Therefore, combining the MIR color-color and SED selection, the AGN fraction in our radio$+$machine-learning identified counterparts of S2COSMOS sources is ($19\pm4$)\%, which is consistent with the AGN fraction in ALMA SMGs of the ALESS  \citep{Wang13} and the AS2UDS samples \citep[$\ls$28\%,][]{Stach19} but lower than a sample of 1.1\,mm-selected ALMA SMGs in the GOODS-South field \citep{Franco18}. 

%
%
\begin{figure}[!t]
\centering
\includegraphics[width=0.46\textwidth]{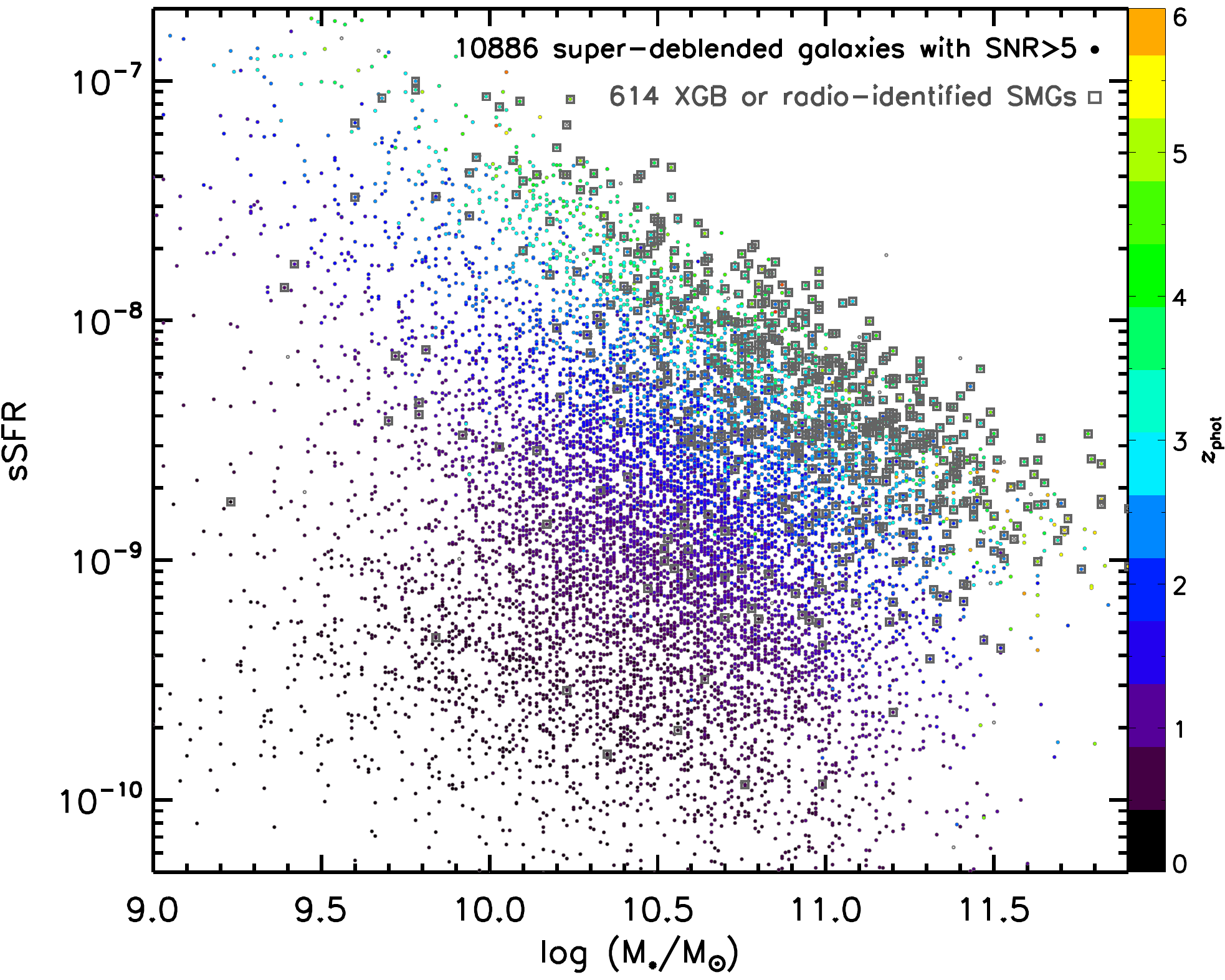}
\caption{Specific star-formation rate (sSFR) versus stellar mass for ``super-deblended'' $K_{\rm s}$ or radio-detected galaxies in the COSMOS field. The SFRs are computed from the integrated 8--1000\,\mm\ infrared luminosities derived from FIR$+$millimeter SED fitting \citep{Jin18}. Stellar mass are from \cite{Laigle16} and \cite{Muzzin13} with a Chabrier IMF \citep{Chabrier03}. We mark our radio$+$machine-learning identified counterparts of SMGs with grey squares. The identified counterparts are the most strongly star-forming and most massive galaxies, compared with the remaining $K_{\rm s}$ or radio-detected field galaxies.}
\label{f:sfr_sm.eps}
\end{figure}

\subsection{Star-formation rate and stellar mass}
SMGs are believed to be massive starburst galaxies with total infrared  luminosity of $L_{\rm IR}\gs$\,10$^{12}$\,$\Lsun$. In this section, we investigate the star-formation rate and stellar mass of our identified counterparts of SCUBA-2 sources in the COSMOS field.

%
%
\begin{figure*}
\centering
\includegraphics[width=0.90\textwidth]{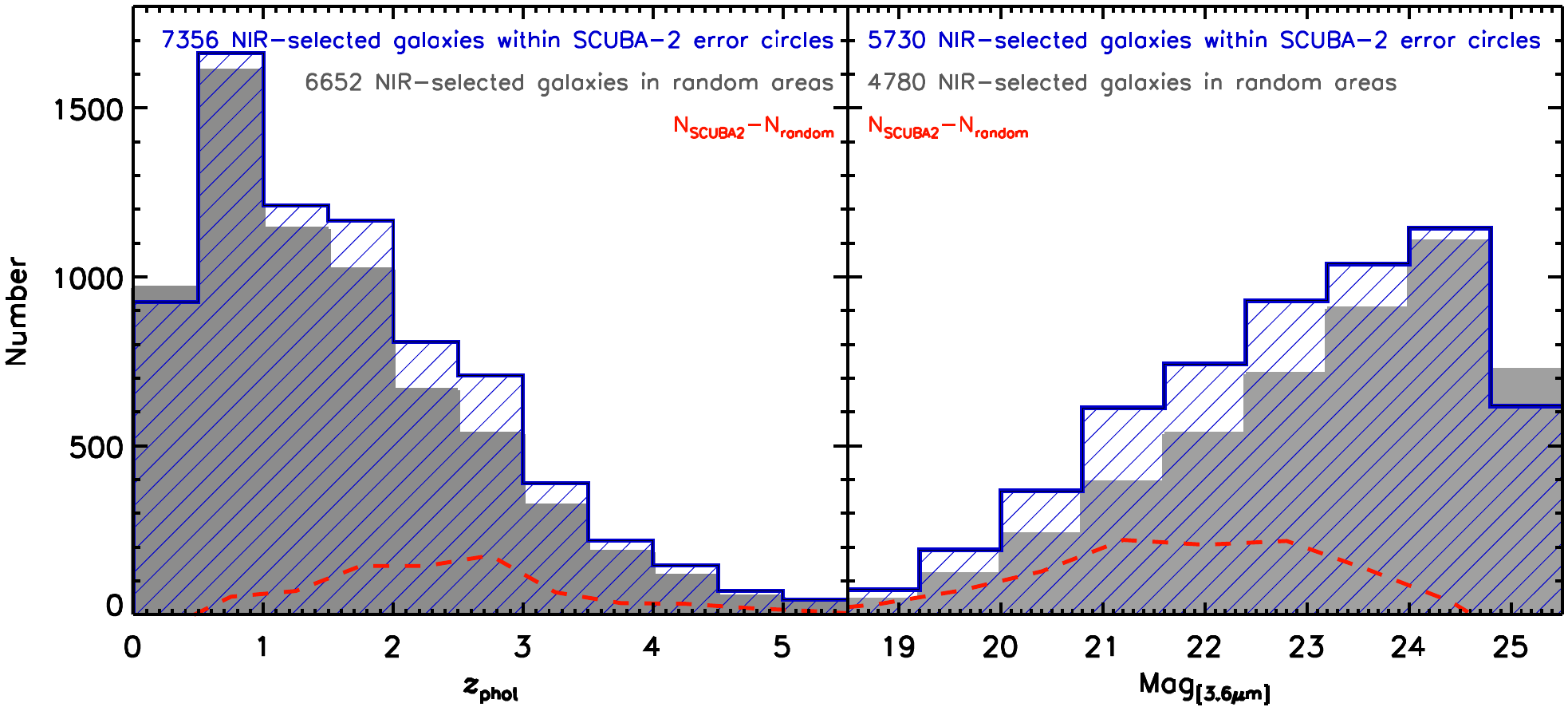}
\caption{The redshift distribution ({\it left}) and 3.6\,\mm\ magnitude ({\it right}) distribution of NIR-selected galaxies within SCUBA-2 error circles (blue shade area), compared to those galaxies within sky (grey area). The galaxies in the right panel also have secure detection at 4.5\,\mm-band and have estimated photometric redshift and absolute $H$-band magnitude from COSMOS2015 catalog, i.e., are qualified for our machine-learning method. We can see an excess in number density within SCUBA-2 error circles, especially at higher redshift and the brighter-end of 3.6\,\mm\ magnitude.}
\label{f:compare_zdistribution.eps}
\end{figure*}

We first match our sample to the FIR to (sub)millimeter photometric catalog from \cite{Jin18}, which is described in Section \ref{s:ONFcatalog}. This catalog is created by using the position of $K_{\rm s}$-band or radio-detected sources and a ``super-deblended'' method developed by \cite{Liu18} to estimate the FIR to (sub)millimeter photometries. The SFR of these ``super-deblended'' sources is estimated by integrating the 8--1000\,\mm\ infrared luminosity from the best-fit FIR$+$mm SED \citep{Jin18}. We show the specific SFR (sSFR), which is defined as the ratio of SFR and stellar mass, as a function of stellar mass for the 10,886 ``super-deblended'' sources with SNR$_{\rm FIR+mm}>5$ in Figure~\ref{f:sfr_sm.eps}. Among these, 614 are our radio$+$machine-learning identified counterparts of S2COSMOS sources. The stellar mass estimates are from the catalogs of \cite{Laigle16} and \cite{Muzzin13} with a Chabrier IMF \citep{Chabrier03}. The photometric redshift of sources is shown by the color scale in Figure~\ref{f:sfr_sm.eps}. 

As shown in Figure~\ref{f:sfr_sm.eps}, our identified counterparts of single-dish submillimeter sources tend to be at higher redshift, while having higher star-formation activity and higher stellar mass, compared with the NIR or radio-selected galaxies with fainter submillimeter emission. These differences support the reliability of our classification. We thus conclude that statistical analyses of SMGs can be undertaken based on this large sample of identified counterparts of submillimeter sources from the panoramic single-dish survey.

\subsection{Excess of number density within SCUBA-2 error circles}\label{s:number_excess}
To study the environment properties of single-dish-detected submillimeter sources, we first compare the number density of NIR-selected galaxies within the SCUBA-2 beam with that in the random areas. These are 1,066/1,147 SCUBA-2 sources lying within the NIR coverage in the COSMOS field. We then randomly offset the 1,066 SCUBA-2 position in Right Ascension or Declination by $9\arcsec$, which is slighter larger than the SCUBA-2 beam, and use the average number of NIR-selected galaxies within these regions as the number density of NIR galaxies in the random areas. We show the distribution of redshift and 3.6\,\mm\ magnitude of NIR galaxies within the SCUBA-2 error circle and within these random areas in Figure~\ref{f:compare_zdistribution.eps}. The number density of NIR galaxies within the SCUBA-2 beam exhibits an excess at high redshift and bright 3.6\,\mm\ magnitudes. The NIR galaxies shown in the right panel of Figure~\ref{f:compare_zdistribution.eps} are qualified for our machine-learning classification, i.e., they have secure detection at 3.6\,\mm\ and 4.5\,\mm-band while having estimated photometric redshift and absolute $H$-band magnitude. The total number of excess NIR galaxies within the SCUBA-2 error circles is 950. In Section \ref{s:identification}, we identified 815 counterparts of SMGs from 5,655 NIR-detected galaxies within these 1066 SCUBA-2 error circles. According to An18, the recall of radio$+$machine-learning method is $\sim$85\% for the AS2UDS SMGs that qualified for machine-learning analyses. Therefore, the excess of NIR galaxies within SCUBA-2 regions can be roughly explained by the contribution from SMGs if we consider the incompleteness of radio$+$machine-learning identification.


\subsection{Properties of ``blank''-SCUBA2 sources}
From the 1,145 SCUBA-2 sources within the VLA 3\,GHz radio survey region or NIR coverage, we find no radio or optical/NIR counterparts for 248 SCUBA-2 sources ($22\pm5$\%). We call these ``blank''-SCUBA-2 sources. We find  only seven radio sources within the error circles of these 248 ``blank''-SCUBA-2 sources and all of them have $p$-value $>$ 0.065, thus are only classified as ``possible'' counterparts. There are 1,960 NIR galaxies within these 248 ``blank''-SCUBA-2 error circles, and 778/1,960 (40\%) of them meet the requirement for the machine-learning analyses, but are classified as non-SMGs. Comparing to   machine-learning identified counterparts of SMGs (Figure~\ref{f:machine_learning.eps}), these 778 NIR galaxies within the ``blank''-SCUBA-2 regions are either fainter in absolute $H$-band or at lower redshift or have a blue NIR color and therefore are classified as ``non-SMGs'' by the XGBoost classifier. 

\begin{figure}
\centering
\includegraphics[width=0.46\textwidth]{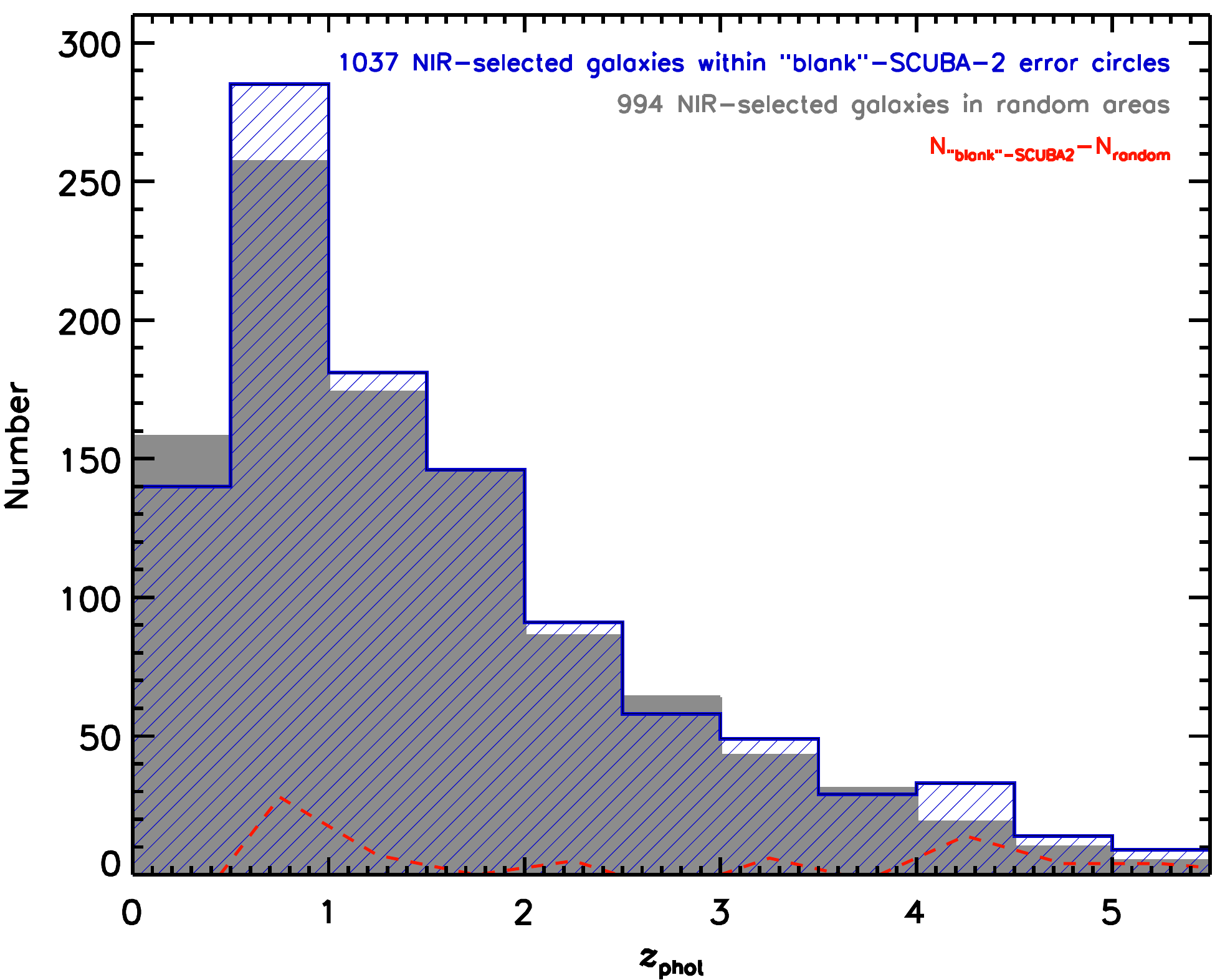}
\caption{The redshift distribution of NIR-selected galaxies within 248 ``blank''-SCUBA-2 error circles, compared with those galaxies within random areas. We can also see an excess of number density within ``blank''-SCUBA-2 error circles, but with a weaker significance compared with that in Figure~\ref{f:compare_zdistribution.eps}. }
\label{f:distribution_sky_blank}
\end{figure}

We also study the environment properties of these ``blank''-SCUBA-2 sources by using the method described in Section \ref{s:number_excess}. There is also an excess of 43 in the number of NIR galaxies that have estimated photometric redshift within the ``blank''-SCUBA-2 sources compared with that in random areas as shown in Figure~\ref{f:distribution_sky_blank}. Therefore, for ``blank''-SCUBA-2 sources, the number density of this excess is 0.17\,beam$^{-1}$, which is weaker than that of all 1066 SCUBA-2 sources with an average density of 0.66\,beam$^{-1}$ as shown in the left panel of Figure~\ref{f:compare_zdistribution.eps}.

We stack the {\it Herschel}/SPIRE maps at the positions of ``blank"-SCUBA-2 sources and detect significant emission with the averaged peak flux densities of $8.3\pm0.4$, $10.5\pm0.6$, and $9.5\pm0.7$\,mJy at 250\,\mm, 350\,\mm, and 500\,\mm, respectively. This suggests that the main explanation for ``blank"-SCUBA-2 sources is the incompleteness of radio$+$machine-learning method. The major causes of the incompleteness is that some SMGs lack radio or NIR counterparts or do not have secure measurements of all the input features required for the machine-learning analyses.

%
%

%
%

%
%
\begin{figure*}
\centering
\includegraphics[width=0.98\textwidth]{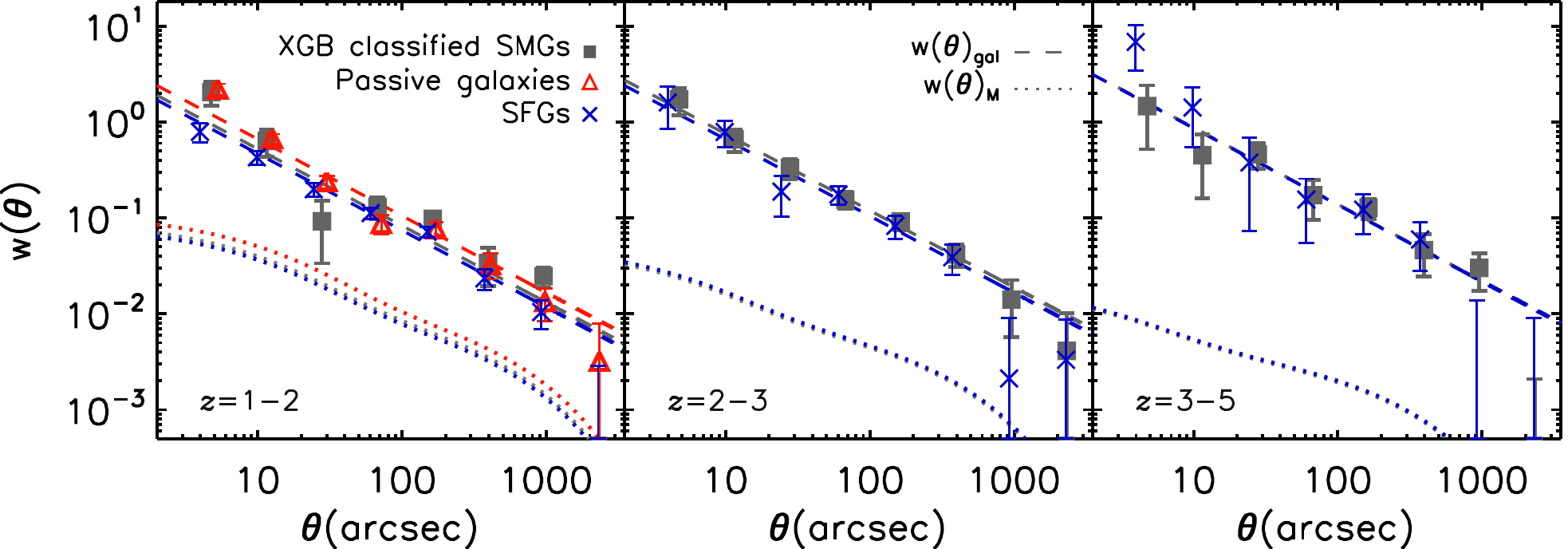}
\caption{Two-point autocorrelation function of machine-learning identified counterparts at $z=$ 1--2, 2--3, and 3--5 in the full COSMOS field, compared with (NUV $-r$)/($r-J$) color-color selected quiescent and star-forming galaxies from the COSMOS2015 catalog \citep{Laigle16}. The stellar mass of galaxies in all three samples are limited with log$_{10}(M_{*}/\Msun)\ge 10.5$. The dashed lines represent the best-fit power-law models, $w(\theta)=A\,\theta^{-0.8}$. The dotted curves represent the autocorrelation function of the total matter. As shown in Appendix \ref{sec:dm_detail}, the redshift distributions of machine-learning identified SMGs, passive galaxies and SFGs are slightly different in each redshift bin. We therefore evaluate the autocorrelation functions of the total matter for the three different samples separately and show them in different colors. In the lowest redshift bin, the passive galaxies show a relatively stronger clustering, compared with machine-learning identified counterparts of SMGs.}
\label{f:autocorrelation.eps}
\end{figure*}

\subsection{SMG clustering}
Spatial clustering is a powerful way to study galaxy evolution and the evolutionary connections between different galaxy populations, since it provides a direct method to constrain the mass of a halo in which galaxies reside \citep[e.g.,][]{Cooray02}.

For the bright SMGs selected at 850\,\mm, the measurements of clustering have suffered from small number statistics because of their low spatial density and small survey area. Previous works resorted to a cross-correlation technique by using other galaxy populations with higher source surface densities \citep{Hickox12, Wilkinson17}. \cite{Chen16b} studied the clustering of SMGs by using a sample of $\sim3000$ OIRTC identified counterparts of SMGs in the UDS field, which includes faint SMGs below the single-dish confusion limit. However, the clustering measurements in \cite{Chen16b} were limited by the moderate survey area of the UDS field. We similarly apply our machine-learning classification to the NIR-selected galaxies in the whole COSMOS field to identify faint SMGs, whose submillimeter emission is fainter than the confusion limit of S2COSMOS surveys. 
Although recent interferometric instruments can detect faint SMGs  \citep[$S_{\rm 850\,\mu m}\ls1.0$\,mJy, e.g., ][]{Franco18, Umehata18}, it is observationally prohibitive to map the full 2\,degree$^2$ COSMOS field with ALMA. By applying our machine-learning method to the whole COSMOS field, we obtain a sample of faint SMG candidates for future interferometric follow-up observations. In total,  356,673 NIR-selected galaxies meet the requirement of the machine-learning classification, i.e., having secure detection at 3.6\,\mm\ and 4.5\,\mm\ and having estimated photometric redshift and absolute $H$-band magnitude in COSMOS2015 catalog \citep{Laigle16}. Among them, 6,877 (2\%) are classified as the likely counterparts of SMGs.  

With this statistically large sample, we investigate the clustering properties of SMGs. We first divide our sample into three redshift bins (Figure~\ref{f:autocorrelation.eps}) to study the evolution of SMGs and compare to clustering results in the literature \citep{Hickox12, Chen16b, Wilkinson17}. We then calculate the two-point autocorrelation function (ACF) $w(\theta)$ for each subsample by using the \cite{Landy93} estimator: 
\begin{eqnarray} \label{e:equation1}
w(\theta)=\frac{\rm (DD-2DR+RR)}{\rm RR},
\end{eqnarray}
 where DD, DR, and RR are the number of data-data, data-random, and random-random galaxy pairs in each $\theta$ bin respectively. The bright stars and bad pixels in the source-detection images of \cite{Laigle16} have been masked out before the clustering analysis. We then generate a random sample within this masked region and with a sample size of  $\sim$1,000 times larger than that of the machine-learning identified counterparts in each redshift bin.

A power-law model $w(\theta)=A\,\theta^{-0.8}$ is assumed for the ACF of galaxies, which is suggested both observationally and theoretically, at the physical scale of $\sim$0.1-10\,$h^{-1}$\,Mpc \citep{Postman98,Zehavi02,Springel05}. However, because our sample is in a field comparable in size to the expected clustering length, the observed ACF needs to be corrected for the integral constraint \citep[IC,][]{Groth77}: 
 \begin{eqnarray} \label{e:equation2}
w(\theta)=A\theta^{-0.8}-{\rm IC}.
\end{eqnarray}
The integral constraint can be numerically estimated \citep[e.g.,][]{Infante94,Roche99, Adelberger05} using the random-random pairs with the following form: 
\begin{eqnarray} \label{e:equation3}
{\rm IC} \approx \int w(\theta) d F_{r},
\end{eqnarray}
where $F_{r}(\theta)$ is the cumulative distribution function of pair angular-separation ($\theta$) estimated from RR counts.

The systematic uncertainties of ACF due to field-to-field variation is estimated by using the Jackknife resampling method \citep[e.g.,][]{Norberg09, Coupon12, Chen16}. In practice, we first divide our sample into $N_{sub}=32$ subsamples and remove one subsample at a time for each Jackknife realization. We then estimate the $w(\theta)_{jk}$ based on each Jackknife sample and repeat this process $N_{\rm sub}$ times. The covariance matrix is derived through the variance of these $w(\theta)_{jk}$. We then fit the observed ACF (Equation~\ref{e:equation2}) by performing a multivariate Gaussian fit on the scales of 8--500$\arcsec$ ($\sim$0.1--7\,$h^{-1}$\,Mpc at $z=2$) where the aforementioned covariance-matrix estimation is used to characterize the correlated uncertainties of the ACF estimator values in each angular-separation bin. The total uncertainties include Poisson noises of number counts and resampling variances, which are propagated into the uncertainty of the normalizing factor $A$ of the ACF model in Equation~\ref{e:equation2}. We show the best-fit results in Figure~\ref{f:autocorrelation.eps}. 

%
%
\begin{figure}[!t]
\centering
\includegraphics[width=0.48\textwidth]{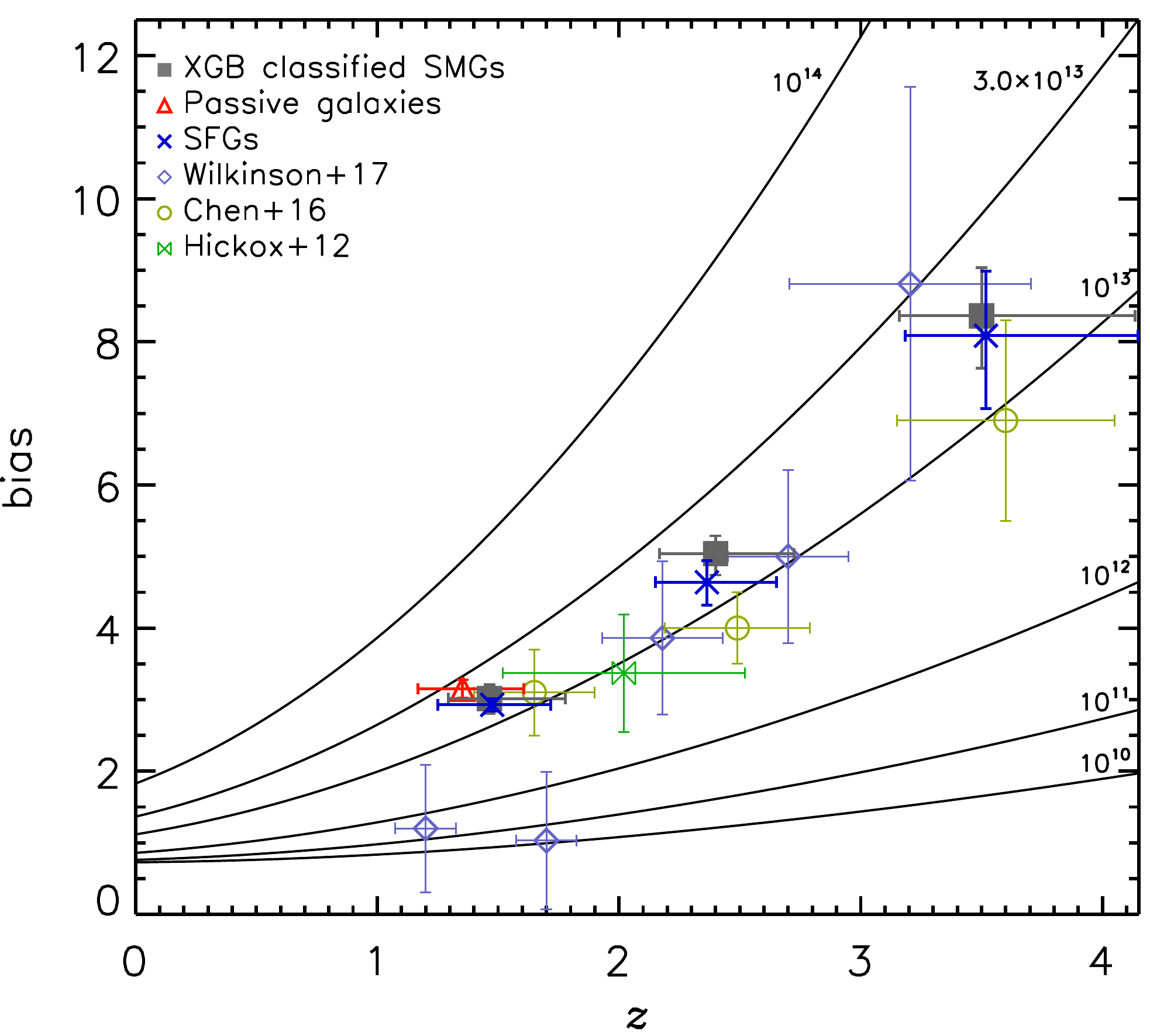}
\caption{Galaxies bias as a function of redshift for machine-learning classified counterparts of SMGs and comparison samples of passive and star-forming galaxies from COSMOS2015 catalog \citep{Laigle16}. We also present the estimated bias of SMGs in literature for comparison. The solid lines show the modelled bias as a function of redshift for various halo masses \citep[labeled by the halo mass at $z=0$ in solar mass,][]{Mo02}. Our measurements show that the typical host halo mass of SMGs with log$_{10}(M_{*}/\Msun)\ge10.5$ is $(1.2\pm0.3)\times10^{13}\,h^{-1}\,\Msun$ at $z>1$.}
\label{f:bais.png}
\end{figure}

To investigate the relation of SMGs to other galaxy populations, we build two comparison samples of star-forming galaxies and passive galaxies using the COSMOS2015 catalog. The star-forming and quiescent galaxies are classified using the location of galaxies in the (NUV $-r$) versus ($r-J$) color-color plane \citep{Laigle16}. Since the clustering properties of galaxies depend on their star-formation rate, stellar mass and redshift \citep[e.g.,][]{Wake11,Mostek13, Coupon15, Wilkinson17,Cochrane18,Lin19}, we match the two comparison samples to our machine-learning classified SMGs in redshift and stellar mass. About 85\% of machine-learning classified SMGs have stellar mass log$_{10}(M_{*}/\Msun)\ge10.5$. Therefore, we limit all three samples with log$_{10}(M_{*}/\Msun)\ge10.5$ in our clustering analyses. 

%
%
\startlongtable
\begin{deluxetable}{ccccc}
\tabletypesize{\scriptsize}
\tablecaption{Results of clustering measurements \label{tab:table2}}
\tablehead{
\colhead{Smaple\tablenotemark{a}} & \colhead{$z_{\rm median}$} & \colhead{$N_{\rm s}$\tablenotemark{b} } & \colhead{$b_{\rm gal}$} & \colhead{M$_{\rm halo}$} \\
& & & & $[10^{13}\,h^{-1}\,\Msun]$
} 
\startdata
   $z=1-2$& & & & \\
   XGB IDs & $1.5^{+0.3}_{-0.2}$ & 2,226 & 3.0$^{+0.2}_{-0.2}$ &1.2$\pm0.3$ \\
   Passive galaxies & $1.4^{+0.3}_{-0.2}$ & 3,478 & 3.2$^{+0.1}_{-0.1}$ &1.7$\pm0.2$ \\ 
   SFGs & $1.5^{+0.3}_{-0.2}$ & 5,337 & 2.9$^{+0.1}_{-0.1}$ &1.0$\pm0.1$\\
   \hline
   $z=2-3$& & & &\\
   XGB IDs & $2.4^{+0.3}_{-0.2}$ & 2,111 & 5.0$^{+0.2}_{-0.3}$ &1.2$\pm0.3$ \\
   SFGs & $2.4^{+0.3}_{-0.2}$ & 1,885 & 4.6$^{+0.3}_{-0.3}$ &1.0$\pm0.3$\\
   \hline
   $z=3-5$& & & & \\
   XGB IDs & $3.5^{+0.6}_{-0.3}$ & 1,072 & 8.4$^{+0.7}_{-0.7}$ &1.3$\pm0.4$\\
   SFGs & $3.5^{+0.6}_{-0.3}$ & 656 & 8.1$^{+0.9}_{-1.0}$ &1.2$\pm0.5$\\
  \enddata
\tablenotetext{a}{The stellar mass limit of all subsamples is log$_{10}(M_{*}/\Msun)\ge10.5$;}
\tablenotetext{b}{The number of sources used in the clustering analyses.}
\end{deluxetable}

For the selection of quiescent galaxies, the $UVJ$/$UrJ$ color-color selection technique has proved to be more reliable at $z<2$ \citep{Williams09, Ilbert13}. In addition, the sample size of quiescent galaxies decreases steeply beyond $z>2$. Therefore, in our analyses, we only include  quiescent galaxies at $1<z<2$. We find that 6.4\% of these red $UrJ$ sources comprise machine-learning identified counterparts of submillimeter sources, which are very likely obscured dusty galaxies. However, the precision of our machine-learning classification being ($83\pm5$)\%, we can not rule out that these sources might be real red passive galaxies. Hence we first keep these overlapped IDs and estimate the ACF for passive galaxies as shown in Figure~\ref{f:autocorrelation.eps}. Then we remove machine-learning IDs from the quiescent sample and estimate their ACF again. We also estimate the effect on ACF of machine-learning classified counterparts of SMGs by removing the IDs that classified as quiescent galaxies. We consider the variances of $w(\theta)$ between these tests as the uncertainties when we calculate the amplitude of $w(\theta)$ for the corresponding sample.

To measure how well galaxies trace the underlying dark matter distribution, we compute the galaxy bias, which is quantified by the relationship
\begin{eqnarray} \label{e:equation4}
b^{2}_{\rm gal}=\frac{w(\theta)}{w(\theta)_{\rm M}},
\end{eqnarray}
where $w(\theta)_{\rm M}$ is the two-point ACF of the total matter, including contributions from both cold dark matter (CDM) and baryons \citep{Desjacques18}.

The $w(\theta)_{\rm M}$ is evaluated based on the small-angle approximation to the projection of isotropic density-fluctuation power spectrum $P(k, z)$ onto a transverse, two-dimensional surface \citep[see][]{Kaiser92, Baugh93, Dodelson03, Loverde08}. Formally, the two-point correlation function is a Hankel transform of the power spectrum,
\begin{equation}
    \label{eq:cdm-angular-corr}
    w(\theta)_{\rm M} = \frac{1}{2 \pi} \int\limits_{0}^{\infty}\!\!
    \int\limits_{0}^{\infty}\! \frac{H(z)}{c} V^2(z) k J_0 \left[ \chi(z) k
    \theta \right] P(k, z) \ dz dk,
\end{equation}
where $H(z)$ is the Hubble parameter, $V(z)$ is the selection function in
redshift $z$ such that $\int_0^{\infty}\! V(z)\ dz = 1$, $\chi(z) =
\int_0^{z}\! c / H(z')\ dz'$ is the radial comoving distance, and $J_0$
denotes the zeroth-order Bessel function of the first kind. We assume a flat $\Lambda$CDM cosmological model with parameters fixed at the
\textit{Planck} 2015 best-fit values \citep{Planck16}. The 
power spectrum $P(k, z)$ in this fiducial cosmology is computed
using the software CAMB \citep{Lewis00,Lewis19}, with the
non-linear evolution of CDM clustering described by the halo model
\citep[see][]{Smith03} as re-calibrated with cosmological $N$-body
simulations by \citet{Takahashi12}. In each redshift bin, the
selection function $V(z)$ is computed from the sources' photometric redshifts ($z_{\rm phot}$), the redshift uncertainties, and their SMG-classification probabilities (the last one only for the machine-learning IDs). The technical details about computing the selection function $V(z)$ are presented in the Appendix \ref{sec:dm_detail}.

We show our measured galaxies bias for machine-learning classified counterparts of SMGs and two comparison samples at three redshifts bins and compare them with the model predictions \citep{Sheth01,Mo02} in Figure~\ref{f:bais.png} and Table~\ref{tab:table2}. We also compare our clustering measurements of these three galaxies populations with the results in the literature \citep{Hickox12, Hartley13,Chen16b, Wilkinson17,Amvrosiadis18, Lin19}. Our results are consistent with the main results in the literature, which suggest that the SMGs resided in high-mass ($(1.2\pm0.3)\times10^{13}\,h^{-1}\,\Msun$) halos. At $z \sim$ 1--2, the passive galaxies show a slightly stronger clustering compared to machine-learning classified counterparts of SMGs. Star-forming galaxies are less clustered compared with the other two populations. This is consistent with the evolutionary scenario that SMGs may be the progenitors of the most massive quiescent galaxies at low redshift.




\section{Conclusion}\label{s:conclusion}
By utilizing the high angular resolution ALMA data and rich ancillary data available in the COSMOS field, we employ our previously developed radio$+$machine-learning method to identify multi-wavelength counterparts of S2COSMOS single-dish-detected submillimeter sources. We then provide a large sample of SMGs with robustly identified radio/optical/NIR counterparts and study the physical and statistical properties of SMGs. Our main conclusions are as follows. 

1. Using the deep VLA 3\,GHz radio data in the COSMOS field, we identify 932 radio counterparts to the S2COSMOS submillimeter sources by adopting a $p$-value cut of $p\le 0.065$. The expected precision of radio identification is 70\% from the self-test of AS2UDS in An18. 

2. We use three ALMA data sets in both COSMOS and UDS fields to build training sets for machine-learning algorithms and compare the performance of machine-learning classifiers trained on these different training sets. The sample constructed from the combined AS2COSMOS and A$^{3}$COSMOS data sets is chosen as the training set for this work because the machine-learning classifiers trained on this sample can best classify SMGs and non-SMGs in the COSMOS field. There are 5,655 NIR galaxies located within SCUBA-2 error circles that meet the requirements of machine-learning classification. Among them, 658 are classified as optical/NIR counterparts of S2COSMOS submillimeter sources by the machine-learning classifier. Combining with the radio identification, we identify 1,222 radio/optical/NIR counterparts to 897 of the 1,145 single-dish-detected submillimeter  sources in the COSMOS field. The identification rate is ($78\pm9$)\% and it increases for bright SCUBA-2 sources. 

3. For the 897 S2COSMOS sources that have at least one radio or machine-learning identified counterpart, ($26\pm5$)\% of them have multiple counterparts. The multiple fraction increases with the flux densities of single-dish submillimeter sources. We estimate that $\sim$27\% of the multiple counterparts within the same SCUBA-2 error circles arise from physically associated galaxies by comparing the difference of their photometric redshift. 

4. We study the physical properties of the 1,222 radio$+$machine-learning identified counterparts to the S2COSMOS submillimeter sources. The redshift distribution of these counterparts peaks at $z=2.3\pm0.1$ and has a redshift range of $z=$\,0.2--5.7, which is consistent with that of ALMA SMGs from AS2UDS and ALESS surveys. The AGN fraction of our radio$+$machine-learning identified counterparts to S2COSMOS sources is ($19\pm4$)\%, which is similar with the AGN fraction in the ALESS and AS2UDS samples. Compared with NIR or radio-selected galaxies in the COSMOS field, our radio$+$machine-learning identified counterparts of S2COSMOS sources have higher star formation rates and higher stellar masses. These results mean that our radio$+$machine-learning identified counterparts constitute a comprehensive and representative sample of SMGs indicated by their physical properties. 

5. We investigate the environment properties of bright SCUBA-2 sources and find a significant excess of NIR galaxies at higher redshifts and brighter NIR magnitudes within SCUBA-2 error circles compared to those within the random fields. We find that the excess of NIR galaxies can be roughly explained by the contribution from SMGs within these regions. 

6. Among the 1,145 S2COSMOS submillimeter sources that lie within the coverage of radio or NIR observations in the COMSOS field, 248 of them do not have any radio or machine-learning identified counterparts. We study the properties of these 248 ``blank"-SCUBA-2 sources and confirm that the main cause of the lack of identified counterpart is the incompleteness of our radio$+$machine-learning method. 

7. We employ our machine-learning technique to the whole COSMOS field and identify 6,877 optical/NIR counterparts of faint SMGs, whose submillimeter emission lies below the confusion limit of our S2COSMOS submillimeter surveys ($S_{\rm 850\mu m}\ls1.5$\,mJy). By using this statistically large sample of SMGs with precisely identified multi-wavelength counterparts, we investigate the clustering properties of this galaxies population. The clustering measurements show that SMGs reside in CDM halos with mass of $\sim(1.2\pm0.3)\times 10^{13}\,h^{-1}\Msun$, which is relatively unchanged across cosmic time. We compare the clustering strength and galaxies bias of SMGs to those of SFGs and passive galaxies at the similar redshift ranges and with the same stellar mass limit of log$_{10}(M_{*}/\Msun)\ge10.5$. We find that at $z=$\,1--2, passive galaxies show a slightly stronger clustering compared with SMGs. Star-forming galaxies are less clustered compared with SMGs at $z=$\,1--5. These results are consistent with the suggested scenario that SMGs may be the progenitors of most massive quiescent galaxies in the low-redshift Universe.

\acknowledgments

FXA acknowledges support from the China Scholarship Council for studying two years at Durham University. All Durham co-authors acknowledge STFC support through grant ST/P000541/1. IRS and BG acknowledge the ERC Advanced Grant DUSTYGAL 321334. JLW acknowledges support from an STFC Ernest Rutherford Fellowship (ST/P004784/2). CM gratefully acknowledges the support of SKA/SA\-RAO/NRF (grant no.~96692). YM acknowledges support from the JSPS grants 17H04831, 17KK0098 and 19H00697.

The James Clerk Maxwell Telescope is operated by the East Asian Observatory on behalf of The National Astronomical Observatory of Japan; Academia Sinica Institute of Astronomy and Astrophysics; the Korea Astronomy and Space Science Institute; the Operation, Maintenance and Upgrading Fund for Astronomical Telescopes and Facility Instruments, budgeted from the Ministry of Finance (MOF) of China and administrated by the Chinese Academy of Sciences (CAS), as well as the National Key R\&D Program of China (No. 2017YFA0402700). Additional funding support is provided by the Science and Technology Facilities Council of the United Kingdom and participating universities in the United Kingdom and Canada (ST/M007634/1, ST/M003019/1, ST/N005856/1). The James Clerk Maxwell Telescope has historically been operated by the Joint Astronomy Centre on behalf of the Science and Technology Facilities Council of the United Kingdom, the National Research Council of Canada and the Netherlands Organization for Scientific Research and data from observations undertaken during this period of operation is used in this manuscript. This research used the facilities of the Canadian Astronomy Data Centre operated by the National Research Council of Canada with the support of the Canadian Space Agency. This paper makes use of the following ALMA data: ADS/JAO.ALMA\#2012.1.00090.S, 2015.1.01528.S,  2016.1.00434.S, and 2016.1.00463.S. ALMA is a partnership of ESO (representing its member states), NSF (USA), and NINS (Japan), together with NRC (Canada), NSC and ASIAA (Taiwan), and KASI (Republic of Korea), in cooperation with the Republic of Chile. The Joint ALMA Observatory is operated by ESO, AUI/NRAO, and NAOJ.

\appendix

\section{Redshift selection function}\label{sec:dm_detail}
Following the discussions of \cite{Chen16b}, in the absence of accurate information about intrinsic redshift and luminosity distributions for the heterogeneous sample \citep{Laigle16}, we may approximate the selection function $V(z)$ by the  PDF of the sample's estimated redshifts in each slice \citep[see also][]{Baugh93}. To do this, it is necessary to account for the redshift uncertainties. In this work, we use a mixture model as our underlying probabilistic assumption, to be detailed as follows.

\begin{figure}[!t]
\centering
\includegraphics[width=0.98\textwidth]{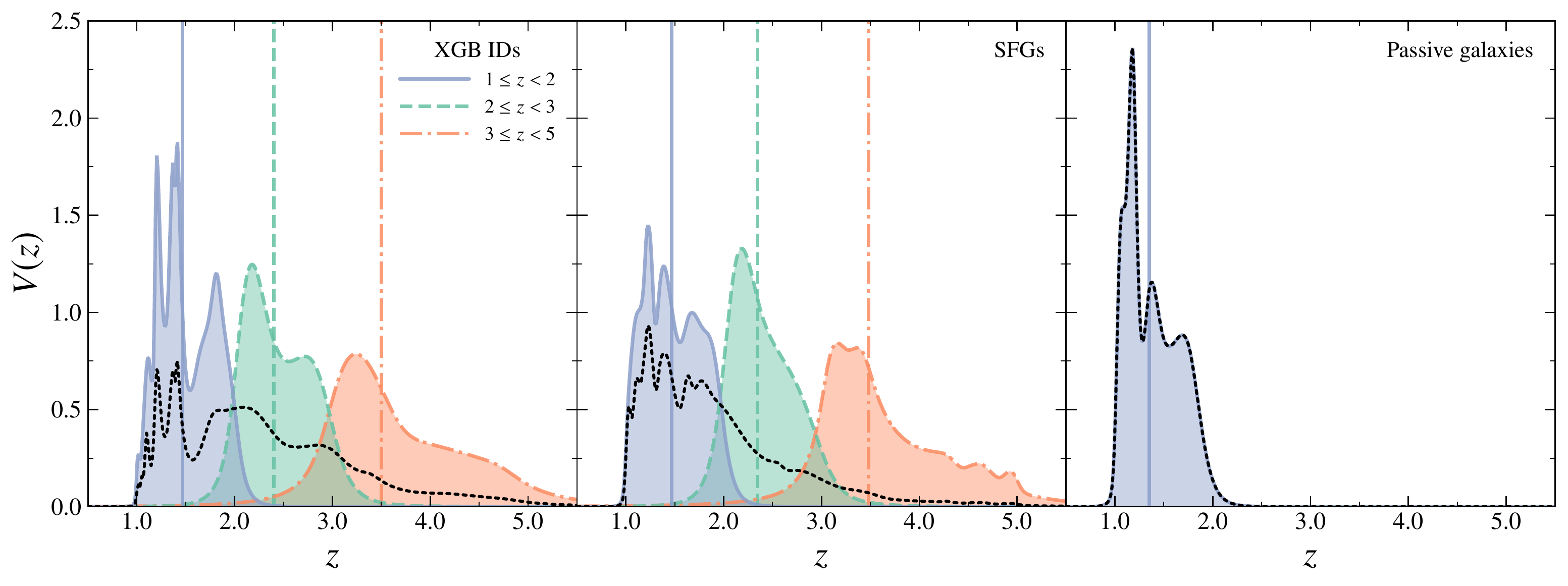}
\caption{Selection functions for subsamples from each redshift bin, with the redshift PDF of the combined sample shown for comparison (dotted lines). Each selection function reflects the aggregated probability density of source redshift from the respective subsample, and is normalized to unity individually. Vertical lines mark the median redshifts of the corresponding subsample (see Table~\ref{tab:table2}).\label{fig:sel-fcn}}
\end{figure}

We denote the redshift (cumulative) distribution function (CDF) of the $i$-th source, as indicated by its $z_{\text{phot}}$ estimation and uncertainty, by $F_i(z)$, and let $p_i$ be the classification probability of the same source.  We consider the mutually-exclusive, equal-probability collection of random events $S_i$, defined as $S_i = \{ \text{the $i$-th source being selected} \}$, such that $\Pr (\bigcup_{i = 0}^{N} S_i) = 1$, where $N$ is the total number of sources in a catalog. Our assumption amounts to $\Pr (S_i) = 1 / N$. Furthermore, we identify each $p_i$ as the conditional probability $\Pr (C\mid S_i)$, where $C$ is the event of inclusion by the classifier. The conditional CDF,
\begin{equation}
    F_{Z}(z) = \Pr \left( \left\{Z < z \right\} \;\middle\vert\; C
    \right),
\end{equation}
is the antiderivative of the selection function $V(z)$ for the whole catalog. To express the selection function based on only a subset (redshift slice) of
the catalog, we denote the index set of the sources in the slice by $\mathcal{J}$ and let $S_{\mathcal{J}} = \bigcup_{i \in \mathcal{J}} S_i$. The CDF specific to (and conditional upon) the slice identified by the indices in $\mathcal{J}$ is

\begin{align}\label{e:equationA1}
    F_{Z}(z; {\mathcal{J}}) &= \Pr \left( \left\{Z < z \right\} \;\middle\vert\; C
    \cap S_{\mathcal{J}} \right) \nonumber \\
    &= \frac{1}{\Pr \left( C \cap S_{\mathcal{J}} \right)} \sum_{i \in \mathcal{J}} \Pr \left(
    \left\{Z < z \right\} \;\middle\vert\; C \cap S_i \right) \Pr \left(C
    \;\middle\vert\; S_i \right) \Pr \left( S_i \right) \nonumber \\
    &= \frac{1}{N \Pr \left( C \cap S_{\mathcal{J}} \right)} \sum_{i \in {\mathcal{J}}} F_i(z) p_i,
 \end{align}
following the definition of conditional probability and the theorem of total probability. The numerical value of $\Pr (C \cap S_{\mathcal{J}})$ is determined by the normalization to unity, $F_{Z}(+\infty; {\mathcal{J}}) = 1$, i.e.,
\begin{equation}
    \Pr \left( C \cap S_{\mathcal{J}} \right) = \frac{1}{N} \sum_{i \in {\mathcal{J}}} p_i.
\end{equation}
It follows that the conditional CDF to be found is a weighted sum of individual components (the mixture model), as in
\begin{equation}
    F_{Z}(z; {\mathcal{J}}) = \frac{\sum_{i \in {\mathcal{J}}} p_i F_i(z)}{\sum_{i \in {\mathcal{J}}} p_i} =
    \sum_{i \in {\mathcal{J}}} w_i F_i(z),
\end{equation}
where $w_i = p_i / (\sum_{i \in {\mathcal{J}}} p_i)$ is the weight. Thus, the selection function for the redshift slice is
\begin{equation}
    \label{eq:vz}
    V(z; {\mathcal{J}}) = \frac{d F_{Z}(z; {\mathcal{J}})}{dz} = \sum_{i \in {\mathcal{J}}} w_i f_i(z),
\end{equation}
where $f_i$ is the PDF characterizing the $z_{\rm{phot}}$ uncertainty of the $i$-th source. In our work, the individual $f_i$'s (see Eq.~[\ref{eq:vz}]) are modeled by Gaussian distributions indicated by $z_{\rm{phot}}$ and the 68.3\% uncertainty bounds. As a result of such uncertainties, the selection function based on a slice can spread outside its redshift cut and overlap with neighboring ones, as can be observed in Figure~\ref{fig:sel-fcn}. Also shown therein is the selection function built from the full catalog, which is comparable to the Figure~3 of \cite{Chen16b}.

%
%
%

%
%
%


\begin{thebibliography}

\bibitem[Adelberger et al.(2005)]{Adelberger05} Adelberger, K.~L., Steidel, C.~C., Pettini, M., et al.\ 2005, \apj, 619, 697 
\bibitem[Amvrosiadis et al.(2018)]{Amvrosiadis18} Amvrosiadis, A., Eales, S.~A., Negrello, M., et al.\ 2018, \mnras, 475, 4939
\bibitem[An et al.(2018)]{An18} An, F.~X., Stach, S.~M., Smail, I., et al.\ 2018, \apj, 862, 101 
\bibitem[Aravena et al.(2016)]{Aravena16} Aravena, M., Decarli, R., Walter, F., et al.\ 2016, \apj, 984, 68 
\bibitem[Aretxaga et al.(2011)]{Aretxaga11} Aretxaga, I., Wilson, G.~W., Aguilar, E., et al.\ 2011, \mnras, 415, 3831
\bibitem[Barger et al.(1998)]{Barger98} Barger, A.\,J., Cowie, L.\,L., Sanders, D.\,B., et al.\ 1998, \nat, 394, 248 
\bibitem[Barger et al.(1999)]{Barger99} Barger, A.\,J., Cowie, L.\,L., Smail, I., et al.\ 1999, \aj, 117, 2656 
\bibitem[Barger et al.(2012)]{Barger12} Barger, A.~J., Wang, W.-H., Cowie, L.~L., et al.\ 2012, \apj, 761, 89 
\bibitem[Baugh \& Efstathiou(1993)]{Baugh93} Baugh, C.~M., \& Efstathiou, G.\ 1993, \mnras, 265, 145
\bibitem[Bertoldi et al.(2007)]{Bertoldi07} Bertoldi, F., Carilli, C., Aravena, M., et al.\ 2007, \apjs, 172, 132
\bibitem[B{\'e}thermin et al.(2012)]{Bethermin12} B{\'e}thermin, M., Le Floc'h, E., Ilbert, O., et al.\ 2012, \aap, 542, A58
\bibitem[Biggs, \& Ivison(2008)]{Biggs08} Biggs, A.~D., \& Ivison, R.~J.\ 2008, \mnras, 385, 893
\bibitem[Bothwell et al.(2013)]{Bothwell13} Bothwell, M.\,S., Smail, I., Chapman, S.\,C., et al.\ 2013, \mnras, 429, 3052
\bibitem[Brusa et al.(2010)]{Brusa10} Brusa, M., Civano, F., Comastri, A., et al.\ 2010, \apj, 716, 348
\bibitem[Cappelluti et al.(2007)]{Cappelluti07} Cappelluti, N., Hasinger, G., Brusa, M., et al.\ 2007, \apjs, 172, 341
\bibitem[Casey et al.(2014)]{Casey14} Casey, C.\,M., Narayanan, D., \& Cooray, A.\ 2014, \physrep, 541, 45 
\bibitem[Chabrier(2003)]{Chabrier03} Chabrier, G.\ 2003, \pasp, 115, 763 
\bibitem[Chapin et al.(2013)]{Chapin13} Chapin, E.~L., Berry, D.~S., Gibb, A.~G., et al.\ 2013, \mnras, 430, 2545 
\bibitem[Chapman et al.(2005)]{Chapman05} Chapman, S.\,C., Blain, A.\,W., Smail, I., \& Ivison, R.\,J.\ 2005, \apj, 622, 772 
\bibitem[Chen \& Guestrin(2016)]{CG16}Chen, T., \& Guestrin, C. 2016, arXiv:1603.02754
\bibitem[Chen et al.(2016a)]{Chen16} Chen, C.-C., Smail, I., Ivison, R.~J., et al.\ 2016, \apj, 820, 82.
\bibitem[Chen et al.(2016b)]{Chen16b} Chen, C.-C., Smail, I., Swinbank, A.~M., et al.\ 2016, \apj, 831, 91 
\bibitem[Civano et al.(2016)]{Civano16} Civano, F., Marchesi, S., Comastri, A., et al.\ 2016, \apj, 819, 62
\bibitem[Clements et al.(2011)]{Clements11} Clements, D.~L., Bendo, G., Pearson, C., et al.\ 2011, \mnras, 411, 373
\bibitem[Cochrane et al.(2018)]{Cochrane18} Cochrane, R.~K., Best, P.~N., Sobral, D., et al.\ 2018, \mnras, 475, 3730
\bibitem[Cooke et al.(2018)]{Cooke18} Cooke, E.~A., Smail, I., Swinbank, A.~M., et al.\ 2018, \apj, 861, 100.
\bibitem[Cooray, \& Sheth(2002)]{Cooray02} Cooray, A., \& Sheth, R.\ 2002, \physrep, 368, 1.
\bibitem[Coppin et al.(2006)]{Coppin06} Coppin, K., Chapin, E.\,L., Mortier, A.\,M.\,J., et al.\ 2006, \mnras, 368, 1621 
\bibitem[Coupon et al.(2012)]{Coupon12} Coupon, J., Kilbinger, M., McCracken, H.~J., et al.\ 2012, \aap, 542, A5.
\bibitem[Coupon et al.(2015)]{Coupon15} Coupon, J., Arnouts, S., van Waerbeke, L., et al.\ 2015, \mnras, 449, 1352
\bibitem[Cowie et al.(2002)]{Cowie02} Cowie, L.~L., Barger, A.~J., \& Kneib, J.-P.\ 2002, \aj, 123, 2197
\bibitem[Cowie et al.(2017)]{Cowie17} Cowie, L.~L., Barger, A.~J., Hsu, L.-Y., et al.\ 2017, \apj, 837, 139
\bibitem[Danielson et al.(2017)]{Danielson17} Danielson, A.\,L.\,R., Swinbank, A.\,M., Smail, I., et al.\ 2017, \apj, 840, 78  
\bibitem[Desjacques et al.(2018)]{Desjacques18} Desjacques, V., Jeong, D., \& Schmidt, F.\ 2018, \physrep, 733, 1
\bibitem[Dodelson(2003)]{Dodelson03} Dodelson, S.\ 2003, Modern Cosmology. San~Diego, CA, United States: Academic Press. ISBN 0-12-219141-2
\bibitem[Donley et al.(2012)]{Donley12} Donley, J.~L., Koekemoer, A.~M., Brusa, M., et al.\ 2012, \apj, 748, 142.
\bibitem[Downes et al.(1986)]{Downes86} Downes, A.\,J.\,B., Peacock, J.\,A., Savage, A., \& Carrie, D.\,R.\ 1986, \mnras, 218, 31 
\bibitem[Dunlop et al.(1989)]{Dunlop89} Dunlop, J.~S., Peacock, J.~A., Savage, A., et al.\ 1989, \mnras, 238, 1171 
\bibitem[Dunlop et al.(2017)]{Dunlop17} Dunlop, J.\,S., McLure, R.\,J., Biggs, A.\,D., et al.\ 2017, \mnras, 466, 861 
\bibitem[Eales et al.(1999)]{Eales99} Eales, S., Lilly, S., Gear, W., et al.\ 1999, \apj, 515, 518
\bibitem[Fawcett\ (2004)]{Fawcett04} Fawcett, T. 2004, Machine Learning, 31, 1
\bibitem[Franco et al.(2018)]{Franco18} Franco, M., Elbaz, D., B{\'e}thermin, M., et al.\ 2018, \aap, 620, A152
\bibitem[Frayer et al.(1998)]{Frayer98} Frayer, D.~T., Seaquist, E.~R., Thuan, T.~X., et al.\ 1998, \apj, 503, 231
\bibitem[Fujimoto et al.(2016)]{Fujimoto16} Fujimoto, S., Ouchi, M., Ono, Y., et al.\ 2016, \apjs, 222, 1
\bibitem[Geach et al.(2017)]{Geach17} Geach, J.~E., Dunlop, J.~S., Halpern, M., et al.\ 2017, \mnras, 465, 1789 
\bibitem[Gear et al.(2000)]{Gear00} Gear, W.~K., Lilly, S.~J., Stevens, J.~A., et al.\ 2000, \mnras, 316, L51
\bibitem[Genzel et al.(2010)]{Genzel10} Genzel, R., Tacconi, L.~J., Gracia-Carpio, J., et al.\ 2010, \mnras, 407, 2091
\bibitem[Groth et al.(1977)]{Groth77} Groth, E.~J., Peebles, P.~J.~E., Seldner, M., et al.\ 1977, Scientific American, 237, 76
\bibitem[Gullberg et al.(2018)]{Gullberg18} Gullberg, B., Swinbank, A.~M., Smail, I., et al.\ 2018, \apj, 859, 12
\bibitem[Hartley et al.(2013)]{Hartley13} Hartley, W.~G., Almaini, O., Mortlock, A., et al.\ 2013, \mnras, 431, 3045
\bibitem[Hasinger et al.(2007)]{Hasinger07} Hasinger, G., Cappelluti, N., Brunner, H., et al.\ 2007, \apjs, 172, 29
\bibitem[Hickox et al.(2012)]{Hickox12} Hickox, R.~C., Wardlow, J.~L., Smail, I., et al.\ 2012, \mnras, 421, 284 
\bibitem[Hodge et al.(2013)]{Hodge13} Hodge, J.\,A., Karim, A., Smail, I., et al.\ 2013, \apj, 768, 91 
\bibitem[Hopkins et al.(2008)]{Hopkins08} Hopkins, P.~F., Hernquist, L., Cox, T.~J., et al.\ 2008, \apjs, 175, 356.
\bibitem[Hughes et al.(1998)]{Hughes98} Hughes, D.\,H., Serjeant, S., Dunlop, J., et al.\ 1998, \nat, 394, 241 
\bibitem[Ilbert et al.(2013)]{Ilbert13} Ilbert, O., McCracken, H.~J., Le F{\`e}vre, O., et al.\ 2013, \aap, 556, A55
\bibitem[Ikarashi et al.(2011)]{Ikarashi11} Ikarashi, S., Kohno, K., Aguirre, J.\,E., et al.\ 2011, \mnras, 415, 3081 
\bibitem[Infante(1994)]{Infante94} Infante, L.\ 1994, \aap, 282, 353
\bibitem[Ishibashi \& Fabian(2016)]{Ishibashi16} Ishibashi, W., \& Fabian, A.~C.\ 2016, \mnras, 463, 1291
\bibitem[Ivison et al.(1998)]{Ivison98} Ivison, R.\,J., Smail, I., Le Borgne, J.-F., et al.\ 1998, \mnras, 298, 583 
\bibitem[Ivison et al.(2002)]{Ivison02} Ivison, R.\,J., Greve, T.\,R., Smail, I., et al.\ 2002, \mnras, 334, 1 
\bibitem[Ivison et al.(2007)]{Ivison07} Ivison, R.\,J., Greve, T.\,R., Dunlop, J.\,S., et al.\ 2007, \mnras, 380, 199 
\bibitem[Jauncey(1968)]{Jauncey68} Jauncey, D.~L.\ 1968, \apj, 152, 647.
\bibitem[Jin et al.(2018)]{Jin18} Jin, S., Daddi, E., Liu, D., et al.\ 2018, \apj, 864, 56 
\bibitem[Kaiser(1992)]{Kaiser92} Kaiser, N.\ 1992, \apj, 388, 272.
\bibitem[Karim et al.(2013)]{Karim13} Karim, A., Swinbank, A.\,M., Hodge, J.\,A., et al.\ 2013, \mnras, 432, 2 
\bibitem[Kohavi et al.(1995)]{Kohavi95} Kohavi, R., et al. 1995, in Proc. of the International Joint Conference on Artificial Intelligence, Vol. 14, 1137

\bibitem[Laigle et al.(2016)]{Laigle16} Laigle, C., McCracken, H.~J., Ilbert, O., et al.\ 2016, \apjs, 224, 24 
\bibitem[Landy \& Szalay(1993)]{Landy93} Landy, S.~D., \& Szalay, A.~S.\ 1993, \apj, 412, 64
\bibitem[Le Floc'h et al.(2009)]{Le09} Le Floc'h, E., Aussel, H., Ilbert, O., et al.\ 2009, \apj, 703, 222
\bibitem[Lewis et al.(2000)]{Lewis00} Lewis, A., Challinor, A., \& Lasenby, A.\ 2000, \apj, 538, 473
\bibitem[Lewis et al.(2019)]{Lewis19} Lewis, A., Vehreschild, A., Mead, A., et~al.\ 2019, CAMB: Code for Anisotropies in the Microwave Background, 1.0.7,  Zenodo
\bibitem[Lin et al.(2019)]{Lin19} Lin, X., Fang, G., Cai, Z.-Y., et al.\ 2019, \apj, 875, 83
\bibitem[Lindner et al.(2011)]{Lindner11} Lindner, R.~R., Baker, A.~J., Omont, A., et al.\ 2011, \apj, 737, 83
\bibitem[Liu et al.(2018)]{Liu18} Liu, D., Daddi, E., Dickinson, M., et al.\ 2018, VizieR Online Data Catalog , J/ApJ/853/172
\bibitem[Liu et al.(2019)]{Liu19} Liu, R.~H., Hill, R., Scott, D., et al.\ 2019, arXiv:1901.09594 
\bibitem[Liu et al.(2019)]{dLiu19} Liu, D., et al.\ 2019, ApJS in press
\bibitem[LoVerde \& Afshordi(2008)]{Loverde08} LoVerde, M., \& Afshordi, N.\ 2008, \prd, 78, 123506.
\bibitem[Lutz et al.(2011)]{Lutz11} Lutz, D., Poglitsch, A., Altieri, B., et al.\ 2011, \aap, 532, A90
\bibitem[Marchesi et al.(2016)]{Marchesi16} Marchesi, S., Civano, F., Elvis, M., et al.\ 2016, \apj, 817, 34.
\bibitem[McMullin et al.(2007)]{McMullin07} McMullin, J.~P., Waters, B., Schiebel, D., Young, W., \& Golap, K.\ 2007, Astronomical Data Analysis Software and Systems XVI, 376, 127 
\bibitem[Micha{\l}owski et al.(2012)]{Michalowski12} Micha{\l}owski, M.~J., Dunlop, J.~S., Ivison, R.~J., et al.\ 2012, \mnras, 426, 1845 
\bibitem[Mo \& White(1996)]{Mo96} Mo, H.~J., \& White, S.~D.~M.\ 1996, \mnras, 282, 347
\bibitem[Mo \& White(2002)]{Mo02} Mo, H.~J., \& White, S.~D.~M.\ 2002, \mnras, 336, 112
\bibitem[Mostek et al.(2013)]{Mostek13} Mostek, N., Coil, A.~L., Cooper, M., et al.\ 2013, \apj, 767, 89
\bibitem[Muzzin et al.(2013)]{Muzzin13} Muzzin, A., Marchesini, D., Stefanon, M., et al.\ 2013, \apj, 777, 18 
\bibitem[Norberg et al.(2009)]{Norberg09} Norberg, P., Baugh, C.~M., Gazta{\~n}aga, E., et al.\ 2009, \mnras, 396, 19
\bibitem[Oke(1974)]{Oke74} Oke, J.\,B.\ 1974, \apjs, 27, 21
\bibitem[Oliver et al.(2010)]{Oliver10} Oliver, S.~J., Wang, L., Smith, A.~J., et al.\ 2010, \aap, 518, L21
\bibitem[Perna et al.(2018)]{Perna18} Perna, M., Sargent, M.~T., Brusa, M., et al.\ 2018, \aap, 619, A90
\bibitem[Planck Collaboration et al.(2016)]{Planck16} Planck Collaboration, Ade, P.~A.~R., Aghanim, N., et al.\ 2016, \aap, 594, A13.
\bibitem[Pope et al.(2006)]{Pope06} Pope, A., Scott, D., Dickinson, M., et al.\ 2006, \mnras, 369, 1185 
\bibitem[Postman et al.(1998)]{Postman98} Postman, M., Lauer, T.~R., Szapudi, I., et al.\ 1998, \apj, 506, 33
\bibitem[Roche \& Eales(1999)]{Roche99} Roche, N., \& Eales, S.~A.\ 1999, \mnras, 307, 703 
\bibitem[Sanders et al.(1988)]{Sanders88} Sanders, D.~B., Soifer, B.~T., Elias, J.~H., et al.\ 1988, \apj, 328, L35
\bibitem[Schinnerer et al.(2010)]{Schinnerer10} Schinnerer, E., Sargent, M.~T., Bondi, M., et al.\ 2010, \apjs, 188, 384
\bibitem[Scott et al.(2002)]{Scott02} Scott, S.\,E., Fox, M.\,J., Dunlop, J.\,S., et al.\ 2002, \mnras, 331, 817 
\bibitem[Scott et al.(2012)]{Scott12} Scott, K.\,S., Wilson, G.\,W., Aretxaga, I., et al.\ 2012, \mnras, 423, 575 
\bibitem[Sheth et al.(2001)]{Sheth01} Sheth, R.~K., Mo, H.~J., \& Tormen, G.\ 2001, \mnras, 323, 1
\bibitem[Simpson et al.(2014)]{Simpson14} Simpson, J.~M., Swinbank, A.~M., Smail, I., et al.\ 2014, \apj, 788, 125 
\bibitem[Simpson et al.(2015a)]{Simpson15a} Simpson, J.\,M., Smail, I., Swinbank, A.\,M., et al.\ 2015, \apj, 807, 128
\bibitem[Simpson et al.(2015b)]{Simpson15b} Simpson, J.\,M., Smail, I., Swinbank, A.\,M., et al.\ 2015, \apj, 815, 81 
\bibitem[Simpson et al.(2017)]{Simpson17} Simpson, J.\,M., Smail, I., Swinbank, A.\,M., et al.\ 2017, \apj, 839, 58 
\bibitem[Simpson et al.(2019)]{Simpson19} Simpson, J.~M., Smail, I., Swinbank, A.~M., et al.\ 2019, \apj, 880, 43
\bibitem[Smail et al.(1997)]{Smail97} Smail, I., Ivison, R.\,J., \& Blain, A.\,W.\ 1997, \apjl, 490, L5 
\bibitem[Smail et al.(2002)]{Smail02} Smail, I., Ivison, R.\,J., Blain, A.\,W., \& Kneib, J.-P.\ 2002, \mnras, 331, 495 
\bibitem[Smith et al.(2003)]{Smith03} Smith, R.~E., Peacock, J.~A., Jenkins, A., et al.\ 2003, \mnras, 341, 1311
\bibitem[Smol{\v c}i{\'c} et al.(2012)]{Smolcic12} Smol{\v c}i{\'c}, V., Aravena, M., Navarrete, F., et al.\ 2012, \aap, 548, A4.
\bibitem[Smol{\v c}i{\'c} et al.(2017a)]{Smolcic17} Smol{\v c}i{\'c}, V., Novak, M., Bondi, M., et al.\ 2017, \aap, 602, A1 
\bibitem[Smol{\v{c}}i{\'c} et al.(2017b)]{Smolcic17b} Smol{\v{c}}i{\'c}, V., Delvecchio, I., Zamorani, G., et al.\ 2017, \aap, 602, A2
\bibitem[Springel et al.(2005)]{Springel05} Springel, V., White, S.~D.~M., Jenkins, A., et al.\ 2005, \nat, 435, 629
\bibitem[Stach et al.(2018)]{Stach18} Stach, S.~M., Smail, I., Swinbank, A.~M., et al.\ 2018, \apj, 860, 161 
\bibitem[Stach et al.(2019)]{Stach19} Stach, S.~M., Dudzevi{\v{c}}i{\={u}}t{\.{e}}, U., Smail, I., et al.\ 2019, \mnras, 487, 4648
\bibitem[Swinbank et al.(2006)]{Swinbank06} Swinbank, A.\,M., Chapman, S.\,C., Smail, I., et al.\ 2006, \mnras, 371, 465 
\bibitem[Swinbank et al.(2014)]{Swinbank14} Swinbank, A.\,M., Simpson, J.\,M., Smail, I., et al.\ 2014, \mnras, 438, 1267 
\bibitem[Swinbank et al.(2015)]{Swinbank15} Swinbank, A.\,M., Dye, S., Nightingale, J.\,W., et al.\ 2015, \apjl, 806, L17
\bibitem[Tacconi et al.(2006)]{Tacconi06} Tacconi, L.~J., Neri, R., Chapman, S.~C., et al.\ 2006, \apj, 640, 228
\bibitem[Takahashi et al.(2012)]{Takahashi12} Takahashi, R., Sato, M., Nishimichi, T., et al.\ 2012, \apj, 761, 152
\bibitem[Targett et al.(2011)]{Targett11} Targett, T.~A., Dunlop, J.~S., McLure, R.~J., et al.\ 2011, \mnras, 412, 295.
\bibitem[Thomson et al.(2014)]{Thomson14} Thomson, A.\,P., Ivison, R.\,J., Simpson, J.\,M., et al.\ 2014, \mnras, 442, 577 
\bibitem[Toft et al.(2014)]{Toft14} Toft, S., Smol{\v c}i{\'c}, V., Magnelli, B., et al.\ 2014, \apj, 782, 68 
\bibitem[Umehata et al.(2018)]{Umehata18} Umehata, H., Hatsukade, B., Smail, I., et al.\ 2018, \pasj, 70, 65

\bibitem[Vapnik (1995)]{Vapnik95} Vapnik, V. N.\ 1995, The Nature of Statistical Learning Theory (Springer)
\bibitem[Wake et al.(2011)]{Wake11} Wake, D.~A., Whitaker, K.~E., Labb{\'e}, I., et al.\ 2011, \apj, 728, 46
\bibitem[Walter et al.(2016)]{Walter16} Walter, F., Decarli, R., Aravena, M., et al.\ 2016, \apj, 984, 67 
\bibitem[Wang et al.(2013)]{Wang13} Wang, S.~X., Brandt, W.~N., Luo, B., et al.\ 2013, \apj, 778, 179.
\bibitem[Wang et al.(2011)]{Wang11} Wang, W.-H., Cowie, L.\,L., Barger, A.\,J., \& Williams, J.\,P.\ 2011, \apjl, 726, L18 
\bibitem[Wang et al.(2017)]{Wang17} Wang, W.-H., Lin, W.-C., Lim, C.-F., et al.\ 2017, \apj, 850, 37 
\bibitem[Wang et al.(2019)]{Wang19} Wang, T., Schreiber, C., Elbaz, D., et al.\ 2019, \nat, 572, 211
\bibitem[Wardlow et al.(2017)]{Wardlow17} Wardlow, J.\,L., Cooray, A., Osage, W., et al.\ 2017, \apj, 837, 12 
\bibitem[Wardlow et al.(2018)]{Wardlow18} Wardlow, J.~L., Simpson, J.~M., Smail, I., et al.\ 2018, \mnras, 479, 3879
\bibitem[Wei{\ss} et al.(2009)]{Weiss09} Wei{\ss}, A., Kov{\'a}cs, A., Coppin, K., et al.\ 2009, \apj, 707, 1211 
\bibitem[Wilkinson et al.(2017)]{Wilkinson17} Wilkinson, A., Almaini, O., Chen, C.-C., et al.\ 2017, \mnras, 464, 1380
\bibitem[Williams et al.(2009)]{Williams09} Williams, R.~J., Quadri, R.~F., Franx, M., et al.\ 2009, \apj, 691, 1879
\bibitem[Williams et al.(2011)]{Williams11} Williams, C.~C., Giavalisco, M., Porciani, C., et al.\ 2011, \apj, 733, 92
\bibitem[Yamamura et al.(2010)]{Yamamura10} Yamamura, I., Makiuti, S., Ikeda, N., et al.\ 2010, VizieR Online Data Catalog, II/298
\bibitem[Yun et al.(2012)]{Yun12} Yun, M.\,S., Scott, K.\,S., Guo, Y., et al.\ 2012, \mnras, 420, 957 
  \bibitem[Zehavi et al.(2002)]{Zehavi02} Zehavi, I., Blanton, M.~R., Frieman, J.~A., et al.\ 2002, \apj, 571, 172
\end{thebibliography}
\end{document}